\documentclass[
    prl, twocolumn, superscriptaddress, amsmath, amssymb,
    aps, floatfix, preprintnumbers
]{revtex4-2}

\usepackage[utf8]{inputenc}
\usepackage{setspace}
\usepackage[dvips]{graphicx}
\usepackage{booktabs}
\usepackage[dvipsnames]{xcolor}
\usepackage{tikz}
\definecolor{CiteBlue}{RGB}{45,52,151}
\usepackage[
    colorlinks=true,
    hypertexnames=false,
    linkcolor=CiteBlue,
    urlcolor=CiteBlue,
    citecolor=CiteBlue
]{hyperref}
\usepackage[version=4]{mhchem}
\usepackage{aas_macros}

\usepackage[capitalise]{cleveref}
\usepackage{siunitx}
\DeclareSIUnit{\year}{yr}

\let\oldsection\section
\usepackage{physics}

\usepackage{bm}
\newcommand{\bb}[1]{\bm{\mathrm{#1}}}

\renewcommand{\Re}{\operatorname{Re}}

\newcommand{\mpid}[1]{\href{https://next-gen.materialsproject.org/materials/mp-#1}{\texttt{mp-#1}}~\cite{MaterialsProject:mp-#1}}

\usepackage[normalem]{ulem}

\newsavebox{\twosubbox}

\begin{document}
%TC:ignore

\title{Unconventional Materials for Light Dark Matter Detection}

\author{Yonit Hochberg}
\affiliation{Racah Institute of Physics, Hebrew University of Jerusalem, Jerusalem 91904, Israel}
\affiliation{Laboratory for Elementary Particle Physics,
 Cornell University, Ithaca, New York 14853, USA}

\author{Dino Novko}
\affiliation{Centre for Advanced Laser Techniques, Institute of Physics, Zagreb 10000, Croatia}
 
\author{Rotem Ovadia}
\affiliation{Racah Institute of Physics, Hebrew University of Jerusalem, Jerusalem 91904, Israel}
\affiliation{Laboratory for Elementary Particle Physics,
 Cornell University, Ithaca, New York 14853, USA}

\author{Antonio Politano}
\affiliation{Department of Physical and Chemical Sciences, University of L'Aquila, L'Aquila 67100, Italy}

\begin{abstract}\ignorespaces{}

We propose the use of several unconventional materials as detectors for dark matter with mass beneath the MeV scale. These include the transition-metal dichalcogenide TiSe$_2$ hosting a low-energy plasmon in the charge-density-wave phase, Sr$_2$RuO$_4$ containing a low-energy acoustic demon mode, and hole-doped diamond with tunable optical and acoustic plasmon frequencies. We perform first-principles density functional theory computations of their loss functions at nonvanishing momenta and establish their reach into light dark matter parameter space. We show that due to intense low-energy plasmon modes --- of different microscopic origin in each --- the reach of detectors based on these materials could surpass existing proposals by several orders of magnitude for both dark matter scattering and absorption on electrons. 
The anisotropic response of these materials, which enables directional detection, renders them exceptionally strong detector candidates, motivating the design and fabrication of future devices.

\end{abstract}

\maketitle

%TC:endignore

The identity of dark matter (DM) in our Universe remains one of the most pressing open questions of modern day physics. The absence of experimental evidence for DM at the electroweak scale, despite years of focus on this regime, has brought lighter DM with mass beneath the GeV scale into the limelight. Indeed, new theoretical frameworks incorporating such light DM have been proposed in recent years (see Ref.~\cite{Asadi:2022njl} for a recent review), along with new experimental avenues with which to detect DM in the sub-GeV and even sub-MeV mass range~\cite{Essig:2011nj,Graham:2012su,Essig:2015cda,Hochberg:2015pha,Hochberg:2015fth,Hochberg:2019cyy,Hochberg:2021ymx,Hochberg:2021yud,Derenzo:2016fse,Hochberg:2016ntt,Hochberg:2017wce,Cavoto:2017otc,Kurinsky:2019pgb,Blanco:2019lrf,Griffin:2020lgd,Simchony:2024kcn,Essig:2022dfa,Das:2022srn,Das:2024jdz,Griffin:2024cew,QROCODILE:2024zmg}. Much progress has been made by interdisciplinary work sitting at the interface of particle physics, condensed matter physics, materials science, and quantum~sensing.

Here, we propose several unconventional materials that can serve as excellent targets for light DM detection. Their common feature is a strong low-energy response in the tens to hundreds of meV energy-deposit regime. Kinematically, nonrelativistic Galactic DM of mass $m_\chi$ has a typical velocity $v \sim 10^{-3}$ in natural units. 
When scattering with a target detector, DM can deposit energy of order its kinetic energy $\sim m_\chi v^2$, corresponding to energy deposits of $\sim 10^{-6} \, m_\chi$, or when absorbed by the detector, it can deposit its entire mass energy $m_\chi$ in the system. 
Consequently, materials with strong responses at energies $\sim 1-100 \, {\rm meV}$ are exceptionally sensitive to DM masses at the keV (for scattering) or meV (for absorption) mass scales.

We highlight a set of low-energy collective excitations in condensed matter systems, which behave as emergent bosonic quasiparticles from far- to mid-infrared energies ($\sim10-100$\,meV) and originate from collective charge dynamics induced by symmetry breaking, nondegenerate multiband response, and charge doping.  Furthermore, the anisotropic response of these excitations enables angular modulation of the DM scattering rate, thereby improving discrimination against isotropic backgrounds and enhancing detection prospects. 
Within this framework, we identify three paradigmatic realizations, each synthesizable at scale, which can serve as excellent targets for light DM detection: ({\it i}) TiSe$_2$---here, a low-energy plasmon mode at few tens of meV appears when a system enters the charge-density-wave (CDW) phase below $T=200$\,K~\cite{Li:2007,Kogar2017,lin2022}; ({\it ii}) Sr$_2$RuO$_4$---here, a gapless acoustic plasmon ({\it i.e.}, a demon mode) emerges as a neutral collective mode generated by out-of-phase charge oscillations between distinct electronic bands~\cite{husain2023}; and ({\it iii}) hole-doped diamond (HDD)---here, a distinct class of infrared plasmons, which are highly tunable with doping, arises due to oscillations of the valence manifold~\cite{bhattacharya2025}. 
We show that the strong low-energy response of these example materials makes them significantly more sensitive than current existing proposals for both DM-electron scattering and absorption. 
In the former two cases, the intrinsic anisotropy of their response enables directional sensitivity through daily modulation, making them promising platforms for DM detection.

%%%%%%%%%%%%%%%%%%%%%%%%%%%%%%%%%%%%%%%%%%%%%%%%%%
\begin{figure*}[t]
    \centering
    \includegraphics[width=0.99\linewidth]{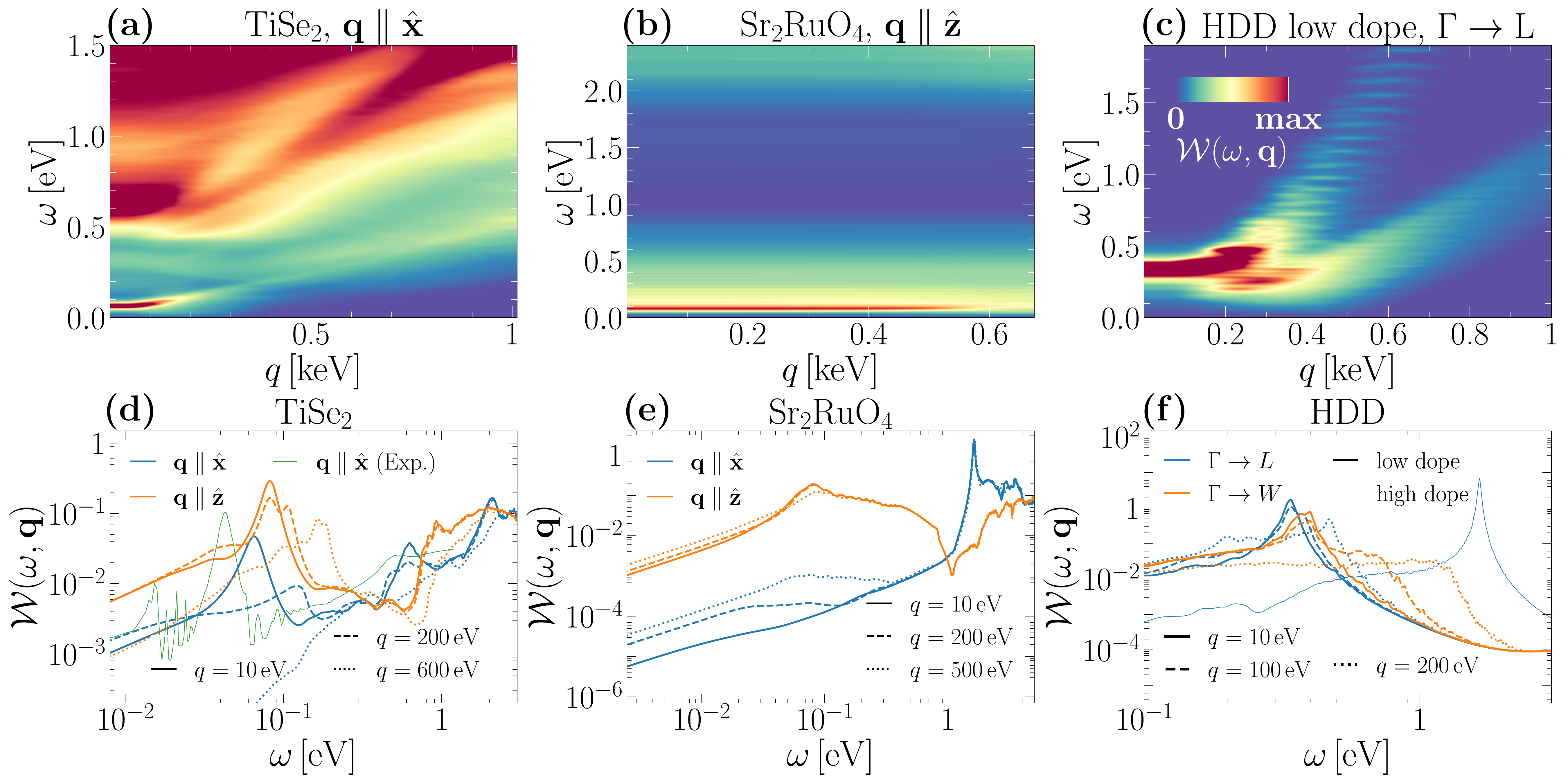}
    \caption{{\bf Loss functions.} Momenta are given in units where $1\,{\rm keV} = 0.507\,{\rm \AA}^{-1}$. Panels {\bf (a)--(c)} display the DFT-calculated loss function along a representative crystal direction, while panels {\bf (d)--(f)} compare responses in two selected directions, shown in blue and orange, across several momenta $q$, indicated by different line styles.
    {\it TiSe$_2$:} panel {\bf (a)} presents the loss function along $\vu{x}$. Panel {\bf (d)} compares responses along $\vu{x}$~(blue) and $\vu{z}$~(orange) for several $q$ values. The green curve delineates the experimentally measured zero-momentum response at $T = 10 \, {\rm K}$ in the $xy$ plane \cite{Li:2007}, to be compared with the solid blue DFT curve at low momenta in the $\vu{x}$ direction.
    {\it Sr$_2$RuO$_4$:} panel {\bf (b)} shows the loss along $\vu{z}$, and panel {\bf (e)} compares the $\vu{x}$ (blue) and $\vu{z}$ (orange) responses.
    {\it HDD:} panel~{\bf (c)} presents the response along $\Gamma \to {\rm L}$ with low doping ($n_h=4.52 \times 10^{19}\,{\rm cm}^{-3}$), while panel {\bf (f)} compares the loss along $\Gamma \to {\rm L}$~(blue) and $\Gamma \to {\rm W}$~(orange). Thick (thin) lines correspond to the low (high) doping $n_h = 4.52 \times 10^{19}\,{\rm cm}^{-3}$ ($n_h = 4.52 \times 10^{21}\,{\rm cm}^{-3}$). 
    }
    \label{fig: loss functions}
\end{figure*}
%%%%%%%%%%%%%%%%%%%%%%%%%%%%%%%%%%%%%%%%%%%%%%%%%%

{\it Unconventional-material loss functions}---We consider spin-independent interactions between DM and the electrons in our unconventional target materials. 
Following the methodology of Refs.~\cite{Hochberg:2021pkt,Boyd:2022tcn,Knapen:2021bwg,Knapen:2021run}, the rate of such interactions [see Eqs.~\eqref{eq:rate} and~\eqref{eq:isotropic-absorption}] is determined by the loss function ${\cal W}(\omega, \mathbf{q}) \equiv {\rm Im}\left\{-1/\epsilon_L(\omega, \mathbf{q})\right\}$, where $\epsilon_L(\omega,\vb{q})=\vu{q}\cdot\overset{\scriptscriptstyle\leftrightarrow}{\epsilon}(\omega,\vb{q})\cdot\vu{q}$ is the longitudinal component of the dielectric tensor for deposited energy $\omega$ and momentum $\vb{q}$ of magnitude $\abs{\vb{q}} = q$. 
The loss encodes the linear response of the material, including many-body effects, and in this Letter is extracted from {\it ab initio} density functional theory~(DFT) calculations at the random-phase-approximation level. 
This gives a simple, calculable and measurable way to assess DM interaction rates in any material of interest.

Here, we present the loss functions and characteristics of our selected materials, which highlight their potential as DM detectors. 
The loss functions of the materials are presented in Fig.~\ref{fig: loss functions}: panels {\bf (a)--(c)} show DFT-computed loss functions in a  representative direction, highlighting the low-energy excitations; panels {\bf (d)--(f)} present the loss along two crystal directions conveying the anisotropy of their response. 
All computed loss functions along with additional computational details are provided in Supplemental Material~(SM)~\cite{supplemental}. 
\nocite{gpaw,yan2011,dressel2002electrodynamics}
We note that the theoretical DFT-based loss functions of Sr$_2$RuO$_4$ and HDD have not been discussed in the literature thus far, and the response for TiSe$_2$ in the whole Brillouin zone (BZ) has been missing~\cite{yin2024}. (For details concerning material synthesis and experimental considerations, see End Matter.)

TiSe$_2$ is a transition-metal dichalcogenide with a trigonal layered structure, which supports a second-order phase transition to CDW at around 200\,K~\cite{disalvo1976}. 
In this CDW phase, TiSe$_2$ undergoes a periodic lattice distortion that breaks the symmetry of the original cell and allows for the gap opening~\cite{Li:2007,lin2022}. 
With this, the concentration of charge carriers and corresponding decay channels drop~\cite{knowles2020}, decreasing the plasmon frequency to $\sim 47 \, {\rm meV}$ and increasing its intensity~\cite{Li:2007,Kogar2017}. 
The interband excitations over the CDW gap form an additional interband plasmon at energies $\sim 600 \, {\rm meV}$. 
We compute the loss function at $T=30\,{\rm K}$ along the $\vu{x}$ and $\vu{z}$ directions.
The response along $\vu{x}$ is shown in Fig.~\ref{fig: loss functions}{\bf (a)}, while Fig.~\ref{fig: loss functions}{\bf (d)} compares responses along both directions at momenta $q=10,200,600\,{\rm eV}$.
As is evident in Fig.~\ref{fig: loss functions}{\bf (a)}, our modeling produces the in-plane bulk plasmon as a sharp peak at energy of $\sim 66 \,  {\rm meV}$, and the interband plasmon as slightly broader and weaker feature at energies $\sim 700 \, {\rm meV}$. Note that the response is anisotropic, in line with the anisotropy of the experimentally reported band structure~\cite{watson2019} and resistivity~\cite{disalvo1976}.

In Fig.~\ref{fig: loss functions}{\bf (d)}, we show our DFT-computed loss function for TiSe$_2$ at various fixed momenta and in different directions, along with optical measurements of the zero-momentum response\,\cite{Li:2007}. 
Since the response is expected to be independent of $q$ (but not of $\vu{q}$) at momenta much smaller than the inverse lattice spacing, one can compare our $q=10 \, {\rm eV}$ DFT-based calculation to the experimental optical data. Two key differences emerge: the measured plasmon frequency is slightly lower in the optical data, and additional phonon excitations appear there that are absent from our DFT result. 
These minor discrepancies are expected, considering the difficulty of simulating the right CDW gap and therefore the screening properties in TiSe$_2$\,\cite{yin2024}. 
We will later show that these differences induce only ${\cal O}(1)$ corrections to the projected DM scattering sensitivity.

The second material we propose is Sr$_2$RuO$_4$, which is a layered perovskite with a body-centered tetragonal structural that falls in the category of strange metals.
Sr$_2$RuO$_4$ exhibits various unconventional features beyond what is expected for a Fermi liquid, such as linear temperature dependence of resistivity and unconventional superconductivity below 1.5\,K~\cite{maeno2024}. 
Recently, a rich electronic response was observed in Sr$_2$RuO$_4$, showing a low-energy acoustic demon mode in addition to a bulk plasmonic mode at 1.6\,eV~\cite{husain2023,schultz2024}.
We compute the loss function of Sr$_2$RuO$_4$ along the $\vu{x}$ and $\vu{z}$ directions, with the loss function along the $\vu{z}$ direction presented in Fig.~\ref{fig: loss functions}{\bf (b)} and the loss functions along both directions at momenta $q=10,200,500\,{\rm eV}$ presented in Fig.~\ref{fig: loss functions}{\bf (e)}.
The low-energy mode is prominent along the $\vu{z}$ direction, manifesting itself as a broad peak in the loss function at energies $\sim 80-800\, {\rm meV}$. 
Furthermore, we find that this mode is suppressed along the $\vu{x}$ axis, producing a large anisotropy in the response.
The bulk plasmon is most prominent along the $\vu{x}$ direction, which exhibits a sharp peak around 1.6\,eV. The obtained anisotropic electronic response of Sr$_2$RuO$_4$ agrees very well with the experimental excitation spectrum probed by electron energy loss spectroscopy~\cite{schultz2024}.

The third candidate is HDD. 
When doped beyond the insulator-metal transition ({\it e.g.}, with boron atoms), diamond exhibits metallic features such as superconductivity at $T_c = 4\, {\rm K}$~\cite{ekimov2004} and strong plasmonic response~\cite{bhattacharya2025}.
Unlike conventional metals, the plasmon frequency in HDD can be tuned by variation of charge carrier density~\cite{bhattacharya2025,pines1956}.
(See also Ref.~\cite{Du:2022dxf} with focus on doped silicon.)
We compute the HDD loss function for two hole dopings~$n_h$ in several directions: at $n_h=4.52\times10^{19}\,{\rm cm}^{-3}$ along $\Gamma\!\to\!{\rm L,X,W,K}$, and at $n_h=4.52\times10^{21}\,{\rm cm}^{-3}$ along $\Gamma\!\to\!{\rm L,X}$.
In Fig.~\ref{fig: loss functions}{\bf (c)} we present the loss function for the lower doping along the $\Gamma \to {\rm L}$ direction and in Fig.~\ref{fig: loss functions}{\bf (f)} along the $\Gamma \to {\rm L,W}$ directions for momenta $q=10,100,200\, {\rm eV}$.
Our computed response consists of an optical plasmon mode centered around $\sim 350$\,meV, along with an acoustic plasmon feature at $\sim 200$\,meV.
The former establishes the relatively strong low-energy response whereas the latter substantially enhances the response at momenta $q > 100 \, {\rm eV}$. 
Both plasmon dispersion relations vary with momenta direction, with larger variations in the acoustic branch. 
We discuss the integrated effect of the acoustic plasmon in our analysis in SM~\cite{supplemental}.
A low-energy electronic response between $100-200$\,meV was recently observed in boron-doped diamond by using electron energy loss and near-field infrared spectroscopies~\cite{bhattacharya2025} and agrees reasonably well with our theoretical results.

Figure~\ref{fig: loss functions}{\bf (f)} compares the response of HDD along two directions and at various momenta. 
Comparing the response in the $\Gamma \to {\rm L}$ direction  at $q = 10 \, {\rm eV}$ for two different dopings, 
we find that the optical plasmon shifts from $\omega_p \sim 0.35\, {\rm eV}$ to $\omega_p \sim 1.6 \, {\rm eV}$ following the typical plasmon to doping density scaling relation $\omega_p \propto n_h^{1/2}$~\cite{pines1956}. This scaling relation is expected to hold up to the metal-insulator critical point below which the plasmon mode is suppressed\,\cite{chen2015,bhattacharya2025}.

{\it Dark matter interaction rates}---For DM-electron scattering via a scalar or vector mediator, the DM-electron interaction rate is given by~\cite{Hochberg:2021pkt, Boyd:2022tcn}
\begin{eqnarray}\label{eq:rate}
	R(t) & = & \frac{1}{\rho_T} \frac{\rho_\chi}{m_\chi} \frac{\pi \bar{\sigma}_e}{\mu^2_{e \chi}} \int \dd^3 \vb{v}\, f(\vb{v}, t) \,  \Gamma(\vb{v})\,,\\[5pt]
	\Gamma(\vb{v}) & = & \int \frac{\dd^3 q}{(2\pi)^3} \abs{{\cal F}(q)}^2 \, \frac{q^2}{2\pi \alpha} \, {\cal W}(\omega_{\vb{q}}(\vb{v}), \vb{q}) \, .
\end{eqnarray}
Here $q = \abs{\vb{q}}$, $\rho_T$ ($\rho_\chi$) is the target (DM) energy density, $\alpha\simeq1/137$ is the fine-structure constant, and $m_\chi$, $m_{\phi}$, $m_e$ are the DM, mediator, and electron masses, respectively. $\mu_{e\chi}$ is the DM-electron reduced mass, and the DM-electron interaction is described by the potential $V(q) = g_e g_\chi / (q^2 + m^2_{\phi})$, where $g_e$ and $g_\chi$ are the mediator couplings to electrons and DM, respectively. Using a reference momentum transfer $q_0 = \alpha m_e$, we define a reference cross section $\bar{\sigma}_e = \mu_{e\chi}^2 \abs{V(q_0)}^2 / \pi$. 
The form factor $\abs{{\cal F}(q)}^2 = \abs{V(q) / V(q_0)}^2$ interpolates between $(q_0/ q)^4$ for light mediators and unity for heavy mediators. 
$f(\vb{v}, t)$ denotes the DM velocity distribution in the lab frame, $\vb{v}$ is the initial DM velocity, with the corresponding energy transfer determined kinematically as $\omega_{\vb{q}}(\vb{v})=\vb{q}\cdot\vb{v}-q^2/(2m_\chi)$. We take $\rho_\chi = 0.3 \, {\rm GeV} / {\rm cm}^{-3}$ and use the Standard Halo Model \cite{Lewin:1995rx} for $f(\vb{v}, t)$ with $v_0 = 220 \,{\rm km}/{\rm s}$, $v_e = 230 \, {\rm km}/{\rm s},$ and $v_{\rm esc} = 540 \, {\rm km} /{\rm s}$.
In addition to accounting for the motion of the Earth in the Galactic plane, the modeling accounts for the rotation of the Earth around its axis, which leads to a daily modulation of the DM scattering rate for detectors with anisotropic responses.

We also consider DM absorption, using the kinetically mixed dark-photon benchmark ${\cal L} \supset -(\kappa / 2) F^{\mu\nu} F'_{\mu\nu}$, where $\kappa$ is the kinetic mixing parameter and $F_{\mu\nu}~(F'_{\mu\nu})$ is the photon (dark-photon) field strength.
The dark-photon absorption rate per unit detector volume per DM particle is given by~\cite{Knapen:2021bwg, Hochberg:2016sqx}
\begin{equation}
	\label{eq:isotropic-absorption}
	\Gamma_{\mathrm{abs}} = \kappa^2 m_\chi
	{\cal W}(m_\chi, m_\chi \vb{v} )
	\,.
\end{equation}
For Galactic light DM, the transferred momentum in an absorption process can be approximated to zero since $|\vb{v}|\sim 10^{-3}$ and consequently $m_\chi|\vb{v}|$ is much smaller than the BZ boundary (typically $\sim {\rm keV}$).

\begin{figure*}[th!]
    \hspace{-0.5cm}\includegraphics[width=1.0\linewidth]{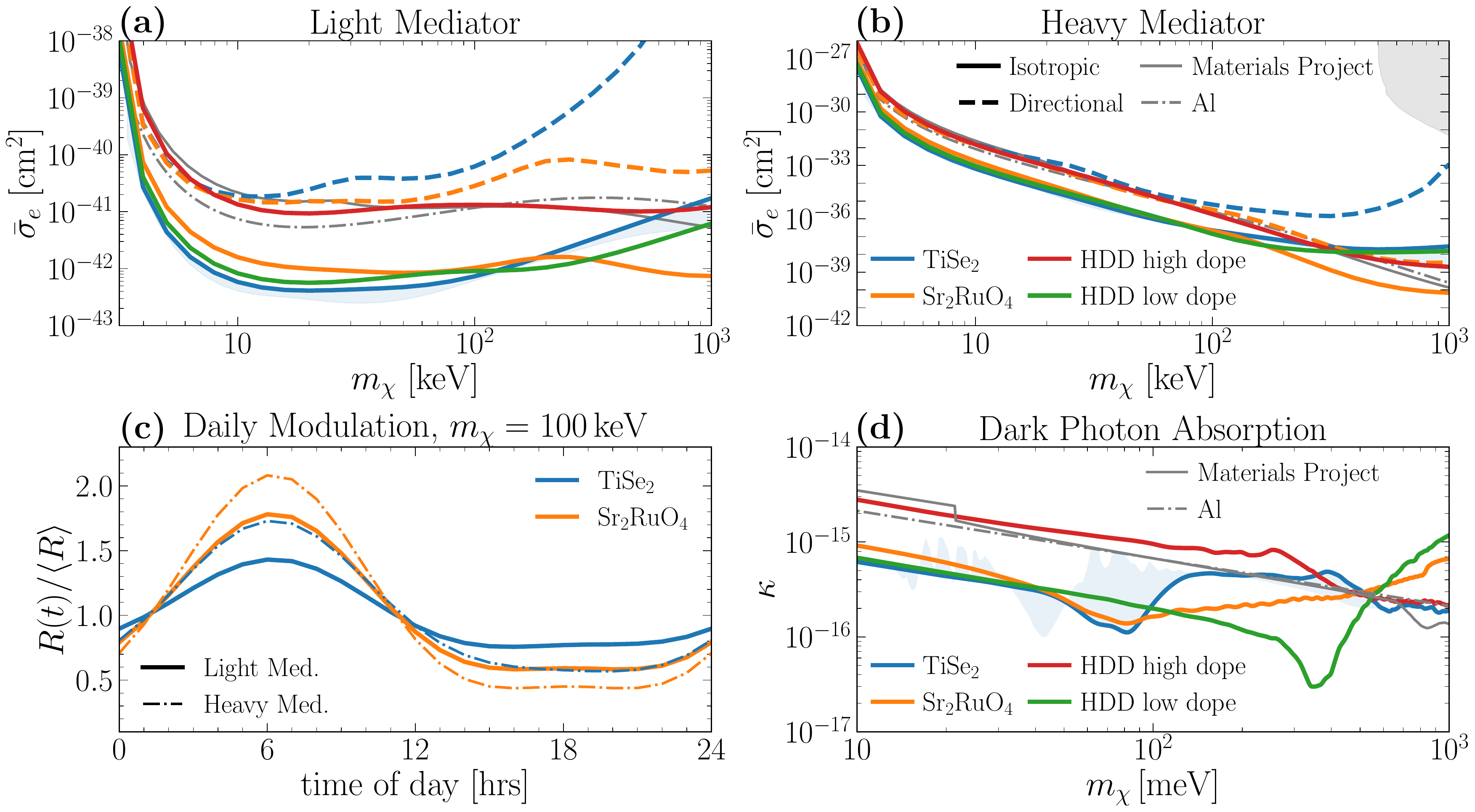}
    
    \caption{{\bf DM reach.} 
    {\bf (a),(b)} {\it Scattering:} projected reach of the candidate materials TiSe$_2$, Sr$_2$RuO$_4$, and HDD for spin-independent DM-electron scattering, assuming a light {\bf (a)} or heavy {\bf (b)} mediator. 
    Solid (dashed thick) colored curves indicate projections  for the isotropic (anisotropic) reach. 
    The high (low) dope HDD curves correspond to hole doping of $n_h = 4.52 \times 10^{21} \, {\rm cm}^{-3}$ ($4.52 \times 10^{19} \, {\rm cm}^{-3}$).  
    The shaded gray region indicates existing terrestrial constraints~\cite{DAMIC:2019dcn,DAMIC-M:2023gxo,SENSEI:2023zdf}.
    {\bf (c)} {\it Daily modulation:} the daily modulation of the DM-target scattering rate resulting from the Earth's rotation around its axis in the Galactic frame, shown for a DM mass $m_\chi = 100\, {\rm keV}$ and a light~(solid curves) or heavy~(dot-dashed curves) mediator. 
    {\bf (d)}~{\it Absorption:} projected reach for detection of a kinetically mixed dark photon.  
    In panels {\bf (a),(b),(d)}, the boundary of the blue shaded region uses experimental optical data of the TiSe$_2$ loss function at zero momentum \cite{Li:2007}, extrapolated when relevant to nonvanishing momenta via a fitting procedure, 
    for comparison with our full DFT results.
    The shaded blue band thus indicates the potential impact of discrepancies between our DFT results and the experimentally measured loss function, which we find to be at most ${\cal O}(1)$.
    In all panels we show results at 95\% CL assuming a kg-yr exposure, no backgrounds, and energy-deposit acceptance in the range $\omega \in [10\,{\rm meV}, 10\, {\rm eV}]$. 
    Thin gray curves delineate the best combined reach of the {\it Materials Project} dataset~\cite{Griffin:2025wew} and superconducting aluminum~\cite{Hochberg:2015pha,Hochberg:2015fth,Hochberg:2021yud} for comparison.}
    \label{fig:reach}
\end{figure*}

{\it Results}---We now present the reach into DM parameter space of the unconventional target materials proposed in this Letter: TiSe$_2$, Sr$_2$RuO$_4$, and HDD. 
We consider the detection prospects for an isotropic signal and for the ability to detect a directional daily modulation signal. 
Details regarding our data handling and statistical analysis are provided in End Matter.
Our projections are done at the 95\% confidence level (CL)~\cite{Feldman:1997qc}, assuming no background, a kg-yr exposure, and a detection sensitivity to energy deposits $\omega \in [10\,{\rm meV}, 10\, {\rm eV}]$.

Our results for DM-electron scattering are shown in Figs.~\ref{fig:reach}{\bf (a)} and~\ref{fig:reach}{\bf (b)}, corresponding to the cases of a light and heavy mediator, respectively.
Solid colored curves indicate the projected isotropic reach.
For TiSe$_2$, the shaded blue band indicates the impact of the uncertainty in the modeling of the material response: the boundary of the blue shaded region uses experimental optical data of the loss
function at zero momentum~\cite{Li:2007}, extrapolated to nonvanishing momenta via the fitting procedure of Ref.~\cite{Griffin:2025wew}, for comparison to our
full DFT results. (Full details of the fitting procedure are provided in SM~\cite{supplemental} for completeness.) We find ${\cal O}(1)$ differences in the scattering rate for this material between the two methods, exhibiting remarkable agreement. 
Dashed curves delineate the anisotropic reach of these materials, indicating their directional discrimination power. 
Here, the ``anisotropic reach'' is defined by the cross section for which the null hypothesis of an isotropic event rate can be rejected at the 95\% CL. 
For HDD we find modulations below the 1\% level and therefore do not present its directional reach.
Note that for the isotropic (anisotropic) reach, we find that a single orientation $\vu{z}$~($\vu{x}$) provides the best reach across the entire presented DM mass range.

We learn that the proposed unconventional materials are capable of probing new regions of light DM parameter space, improving the reach well-beyond established benchmarks such as superconducting aluminum~\cite{Hochberg:2015pha,Hochberg:2015fth,Hochberg:2021yud} and even the best materials found in the large material science dataset of the {\it Materials Project}~\cite{Griffin:2025wew} (where, {\it e.g.}, LiMoO$_2$ [\mpid{19338}] and C [\mpid{569304}] have the best reach in the mass range considered here for scattering).
TiSe$_2$ in particular outperforms both aluminum and the {\it Materials Project} database by up to 2 to 3 orders of magnitude, respectively, over several DM mass decades, with strong directional detection prospects.

The daily modulation of the scattering rate in Sr$_2$RuO$_4$ and TiSe$_2$ is shown in Fig.~\ref{fig:reach}{\bf (c)}, illustrated for a $m_\chi = 100\,{\rm keV}$ DM signal for light~({\it solid}) and heavy~({\it dot-dashed}) mediators. 
The directionally dependent response is computed using a symmetry-aware angular interpolation of the DFT dielectric data, incorporating all crystallographically equivalent directions. 
The modulation arises from the Earth's daily rotation, which continuously changes the DM wind orientation relative to the crystal axes.
Sr$_2$RuO$_4$ and TiSe$_2$ exhibit especially large daily modulation signals due to their anisotropic low-energy plasmon features that are present along specific axes and suppressed along others.
Unsurprisingly, the heavy mediator model exhibits larger modulations since it probes higher momentum transfers, where interaxis differences in the response are more pronounced. 
For instance, the in-plane and out-of-plane plasmonic responses in Sr$_2$RuO$_4$ are strongly disparate, and hence the corresponding modulation is the largest among the considered systems.
Additionally, we find that despite substantial differences in the response across HDD doping levels, the shape and amplitude of the modulation remain nearly identical since the anisotropy of the response is preserved within the considered doping range.

Figure \ref{fig:reach}{\bf (d)} shows the prospects for detection of absorption of kinetically mixed dark photon DM.
The shaded blue region delineates the difference between the projection of our DFT low momenta results for TiSe$_2$ and the use of fitted optical data \cite{Li:2007}. The qualitative ${\cal O}(1)$ differences between the two absorption rates for this material are attributed to the known limitations of our DFT modeling, as discussed earlier. We learn that TiSe$_2$, Sr$_2$RuO$_4$, and HDD at low doping
surpass the reach of the best candidates from the {\it Materials Project}~\cite{Griffin:2025wew} and of superconducting aluminum~\cite{Hochberg:2015pha,Hochberg:2015fth,Hochberg:2021yud}, consolidating their standing as exceptional candidate targets for DM searches across many mass scales.

{\it Outlook}---We have identified three classes of unconventional materials with low-energy collective excitations---transition-metal dichalcogenides with a CDW phase, multilayered perovskites, and doped multiband semiconductors---that provide enhanced sensitivity to dark matter scattering (absorption) at ${\rm keV}$-${\rm MeV}$ ($\rm{meV}$-${\rm eV}$) masses. 
Focusing on candidate materials in each class—TiSe$_2$, Sr$_2$RuO$_4$, and HDD—we demonstrated that detectors based on these materials could surpass existing proposals by several orders of magnitude in isotropic searches, with TiSe$_2$ and Sr$_2$RuO$_4$ additionally exhibiting strong directional sensitivity.
Our Letter utilizes the dielectric function formalism~\cite{Hochberg:2021pkt} along with state-of-the-art material science in order to identify materials with incredible prospects for DM detection and to establish their reach.

The proposed materials are experimentally accessible, with TiSe$_2$ in particular synthesizable on a large scale: milligram-size crystals are readily synthesized, and kg-scale targets are achievable on the timescale of months of furnace operation at minimal cost.     
The strong, anisotropic low-energy responses of the proposed materials open new avenues for designing detectors that improve sensitivity to light DM and enable a robust mechanism for background discrimination. 
Existing low-threshold readout technologies employed in other DM detection experiments can be applied to our proposed targets, enabling the construction of an end-to-end light DM experiment based on these materials.

Our results motivate the exploration of these and similar materials as the constituents of next-generation DM detectors, such as transition-metal dichalcogenides with low-energy plasmons~\cite{gjerding2017} and cuprate perovskites~\cite{nag2020}.
Low-energy deposits in these materials should allow the detection of DM-nucleon interactions as well through phonon production~\cite{Griffin:2024cew}, as will be explored in future work. 
More broadly, our Letter emphasizes that ${\cal O}({\rm meV})$-scale collective modes in solid-state systems have yet to be comprehensively mapped out in the context of DM. 
Fundamental physics will gain much from an interdisciplinary cohesive effort that classifies and identifies relevant materials targeting light DM~detection.

%TC:ignore
\bigskip
%%%%%%%%%%%%%%%%%%%%%%%%%%%%%%%%%
\begin{acknowledgments}
{\it Acknowledgments}---We are grateful to the Italian Embassy in Israel for initiating the ``Italy-Israel Bilateral Online Forum on Quantum Materials,'' which led to this joint work. We thank Ben Lehmann for comments on the manuscript. 
The work of Y.~H. is supported by the Israel Science Foundation (Grant No. 1818/22), by the Binational Science Foundation (Grants No. 2018140 and No. 2022287) and by an ERC STG grant (``Light-Dark,'' Grant No. 101040019). 
D.~N. acknowledges financial support from the project ``Podizanje znanstvene izvrsnosti Centra za napredne laserske tehnike (CALTboost)" financed by the European Union through the National Recovery and Resilience Plan 2021--2026 (NRPP) and European Regional Development Fund for the project ``Materials for clean energy, advanced sensors and quantum technologies'' (Grant No. PK.1.1.10.0002).
R.~O. acknowledges partial support from the ERC STG Grant ``Light-Dark" (Grant No. 101040019), BSF Travel Grant No. 3083000028, the Milner Fellowship, the Azrieli Fellowship, and the gracious hospitality of Cornell University.
The work by A.~P. was funded by the European Commission-Next Generation EU, Mission 4 Component C2, Investment 1.1, under the Ministry of University and Research (MUR) of Italy PRIN 2022 (CUP: E53D23001750006, Grant No. 2022LFWJBR, acronym PLANET) and PRIN PNRR (CUP: E53D23018280001, Grant No. P20223LXTA, acronym ENTANGLE) projects. This project has received funding from the European Research Council (ERC) under the European Union’s Horizon Europe research and innovation programme (Grant Agreement No. 101040019).
Views and opinions expressed are however those of the authors only and do not necessarily reflect those of the European Union. The European Union cannot be held responsible for them. 
\end{acknowledgments}

\bibliography{references}

%apsrev4-2.bst 2019-01-14 (MD) hand-edited version of apsrev4-1.bst
%Control: key (0)
%Control: author (8) initials jnrlst
%Control: editor formatted (1) identically to author
%Control: production of article title (0) allowed
%Control: page (0) single
%Control: year (1) truncated
%Control: production of eprint (0) enabled
\begin{thebibliography}{61}%
\makeatletter
\providecommand \@ifxundefined [1]{%
 \@ifx{#1\undefined}
}%
\providecommand \@ifnum [1]{%
 \ifnum #1\expandafter \@firstoftwo
 \else \expandafter \@secondoftwo
 \fi
}%
\providecommand \@ifx [1]{%
 \ifx #1\expandafter \@firstoftwo
 \else \expandafter \@secondoftwo
 \fi
}%
\providecommand \natexlab [1]{#1}%
\providecommand \enquote  [1]{``#1''}%
\providecommand \bibnamefont  [1]{#1}%
\providecommand \bibfnamefont [1]{#1}%
\providecommand \citenamefont [1]{#1}%
\providecommand \href@noop [0]{\@secondoftwo}%
\providecommand \href [0]{\begingroup \@sanitize@url \@href}%
\providecommand \@href[1]{\@@startlink{#1}\@@href}%
\providecommand \@@href[1]{\endgroup#1\@@endlink}%
\providecommand \@sanitize@url [0]{\catcode `\\12\catcode `\$12\catcode
  `\&12\catcode `\#12\catcode `\^12\catcode `\_12\catcode `\%12\relax}%
\providecommand \@@startlink[1]{}%
\providecommand \@@endlink[0]{}%
\providecommand \url  [0]{\begingroup\@sanitize@url \@url }%
\providecommand \@url [1]{\endgroup\@href {#1}{\urlprefix }}%
\providecommand \urlprefix  [0]{URL }%
\providecommand \Eprint [0]{\href }%
\providecommand \doibase [0]{https://doi.org/}%
\providecommand \selectlanguage [0]{\@gobble}%
\providecommand \bibinfo  [0]{\@secondoftwo}%
\providecommand \bibfield  [0]{\@secondoftwo}%
\providecommand \translation [1]{[#1]}%
\providecommand \BibitemOpen [0]{}%
\providecommand \bibitemStop [0]{}%
\providecommand \bibitemNoStop [0]{.\EOS\space}%
\providecommand \EOS [0]{\spacefactor3000\relax}%
\providecommand \BibitemShut  [1]{\csname bibitem#1\endcsname}%
\let\auto@bib@innerbib\@empty
%</preamble>
\bibitem [{\citenamefont {Asadi}\ \emph {et~al.}(2022)\citenamefont {Asadi}
  \emph {et~al.}}]{Asadi:2022njl}%
  \BibitemOpen
  \bibfield  {author} {\bibinfo {author} {\bibfnamefont {P.}~\bibnamefont
  {Asadi}} \emph {et~al.},\ }\href@noop {} {\bibinfo {title} {{Early-Universe
  Model Building}}} (\bibinfo {year} {2022}),\ \Eprint
  {https://arxiv.org/abs/2203.06680} {arXiv:2203.06680 [hep-ph]} \BibitemShut
  {NoStop}%
\bibitem [{\citenamefont {Essig}\ \emph {et~al.}(2012)\citenamefont {Essig},
  \citenamefont {Mardon},\ and\ \citenamefont {Volansky}}]{Essig:2011nj}%
  \BibitemOpen
  \bibfield  {author} {\bibinfo {author} {\bibfnamefont {R.}~\bibnamefont
  {Essig}}, \bibinfo {author} {\bibfnamefont {J.}~\bibnamefont {Mardon}},\ and\
  \bibinfo {author} {\bibfnamefont {T.}~\bibnamefont {Volansky}},\ }\bibfield
  {title} {\bibinfo {title} {{Direct Detection of Sub-GeV Dark Matter}},\
  }\href {https://doi.org/10.1103/PhysRevD.85.076007} {\bibfield  {journal}
  {\bibinfo  {journal} {Phys. Rev. D}\ }\textbf {\bibinfo {volume} {85}},\
  \bibinfo {pages} {076007} (\bibinfo {year} {2012})},\ \Eprint
  {https://arxiv.org/abs/1108.5383} {arXiv:1108.5383 [hep-ph]} \BibitemShut
  {NoStop}%
\bibitem [{\citenamefont {Graham}\ \emph {et~al.}(2012)\citenamefont {Graham},
  \citenamefont {Kaplan}, \citenamefont {Rajendran},\ and\ \citenamefont
  {Walters}}]{Graham:2012su}%
  \BibitemOpen
  \bibfield  {author} {\bibinfo {author} {\bibfnamefont {P.~W.}\ \bibnamefont
  {Graham}}, \bibinfo {author} {\bibfnamefont {D.~E.}\ \bibnamefont {Kaplan}},
  \bibinfo {author} {\bibfnamefont {S.}~\bibnamefont {Rajendran}},\ and\
  \bibinfo {author} {\bibfnamefont {M.~T.}\ \bibnamefont {Walters}},\
  }\bibfield  {title} {\bibinfo {title} {{Semiconductor Probes of Light Dark
  Matter}},\ }\href {https://doi.org/10.1016/j.dark.2012.09.001} {\bibfield
  {journal} {\bibinfo  {journal} {Phys. Dark Univ.}\ }\textbf {\bibinfo
  {volume} {1}},\ \bibinfo {pages} {32} (\bibinfo {year} {2012})},\ \Eprint
  {https://arxiv.org/abs/1203.2531} {arXiv:1203.2531 [hep-ph]} \BibitemShut
  {NoStop}%
\bibitem [{\citenamefont {Essig}\ \emph {et~al.}(2016)\citenamefont {Essig},
  \citenamefont {Fernandez-Serra}, \citenamefont {Mardon}, \citenamefont
  {Soto}, \citenamefont {Volansky},\ and\ \citenamefont {Yu}}]{Essig:2015cda}%
  \BibitemOpen
  \bibfield  {author} {\bibinfo {author} {\bibfnamefont {R.}~\bibnamefont
  {Essig}}, \bibinfo {author} {\bibfnamefont {M.}~\bibnamefont
  {Fernandez-Serra}}, \bibinfo {author} {\bibfnamefont {J.}~\bibnamefont
  {Mardon}}, \bibinfo {author} {\bibfnamefont {A.}~\bibnamefont {Soto}},
  \bibinfo {author} {\bibfnamefont {T.}~\bibnamefont {Volansky}},\ and\
  \bibinfo {author} {\bibfnamefont {T.-T.}\ \bibnamefont {Yu}},\ }\bibfield
  {title} {\bibinfo {title} {{Direct Detection of sub-GeV Dark Matter with
  Semiconductor Targets}},\ }\href {https://doi.org/10.1007/JHEP05(2016)046}
  {\bibfield  {journal} {\bibinfo  {journal} {JHEP}\ }\textbf {\bibinfo
  {volume} {05}},\ \bibinfo {pages} {046}},\ \Eprint
  {https://arxiv.org/abs/1509.01598} {arXiv:1509.01598 [hep-ph]} \BibitemShut
  {NoStop}%
\bibitem [{\citenamefont {Hochberg}\ \emph
  {et~al.}(2016{\natexlab{a}})\citenamefont {Hochberg}, \citenamefont {Zhao},\
  and\ \citenamefont {Zurek}}]{Hochberg:2015pha}%
  \BibitemOpen
  \bibfield  {author} {\bibinfo {author} {\bibfnamefont {Y.}~\bibnamefont
  {Hochberg}}, \bibinfo {author} {\bibfnamefont {Y.}~\bibnamefont {Zhao}},\
  and\ \bibinfo {author} {\bibfnamefont {K.~M.}\ \bibnamefont {Zurek}},\
  }\bibfield  {title} {\bibinfo {title} {{Superconducting Detectors for
  Superlight Dark Matter}},\ }\href
  {https://doi.org/10.1103/PhysRevLett.116.011301} {\bibfield  {journal}
  {\bibinfo  {journal} {Phys. Rev. Lett.}\ }\textbf {\bibinfo {volume} {116}},\
  \bibinfo {pages} {011301} (\bibinfo {year} {2016}{\natexlab{a}})},\ \Eprint
  {https://arxiv.org/abs/1504.07237} {arXiv:1504.07237 [hep-ph]} \BibitemShut
  {NoStop}%
\bibitem [{\citenamefont {Hochberg}\ \emph
  {et~al.}(2016{\natexlab{b}})\citenamefont {Hochberg}, \citenamefont {Pyle},
  \citenamefont {Zhao},\ and\ \citenamefont {Zurek}}]{Hochberg:2015fth}%
  \BibitemOpen
  \bibfield  {author} {\bibinfo {author} {\bibfnamefont {Y.}~\bibnamefont
  {Hochberg}}, \bibinfo {author} {\bibfnamefont {M.}~\bibnamefont {Pyle}},
  \bibinfo {author} {\bibfnamefont {Y.}~\bibnamefont {Zhao}},\ and\ \bibinfo
  {author} {\bibfnamefont {K.~M.}\ \bibnamefont {Zurek}},\ }\bibfield  {title}
  {\bibinfo {title} {{Detecting Superlight Dark Matter with Fermi-Degenerate
  Materials}},\ }\href {https://doi.org/10.1007/JHEP08(2016)057} {\bibfield
  {journal} {\bibinfo  {journal} {JHEP}\ }\textbf {\bibinfo {volume} {08}},\
  \bibinfo {pages} {057}},\ \Eprint {https://arxiv.org/abs/1512.04533}
  {arXiv:1512.04533 [hep-ph]} \BibitemShut {NoStop}%
\bibitem [{\citenamefont {Hochberg}\ \emph {et~al.}(2019)\citenamefont
  {Hochberg}, \citenamefont {Charaev}, \citenamefont {Nam}, \citenamefont
  {Verma}, \citenamefont {Colangelo},\ and\ \citenamefont
  {Berggren}}]{Hochberg:2019cyy}%
  \BibitemOpen
  \bibfield  {author} {\bibinfo {author} {\bibfnamefont {Y.}~\bibnamefont
  {Hochberg}}, \bibinfo {author} {\bibfnamefont {I.}~\bibnamefont {Charaev}},
  \bibinfo {author} {\bibfnamefont {S.-W.}\ \bibnamefont {Nam}}, \bibinfo
  {author} {\bibfnamefont {V.}~\bibnamefont {Verma}}, \bibinfo {author}
  {\bibfnamefont {M.}~\bibnamefont {Colangelo}},\ and\ \bibinfo {author}
  {\bibfnamefont {K.~K.}\ \bibnamefont {Berggren}},\ }\bibfield  {title}
  {\bibinfo {title} {{Detecting Sub-GeV Dark Matter with Superconducting
  Nanowires}},\ }\href {https://doi.org/10.1103/PhysRevLett.123.151802}
  {\bibfield  {journal} {\bibinfo  {journal} {Phys. Rev. Lett.}\ }\textbf
  {\bibinfo {volume} {123}},\ \bibinfo {pages} {151802} (\bibinfo {year}
  {2019})},\ \Eprint {https://arxiv.org/abs/1903.05101} {arXiv:1903.05101
  [hep-ph]} \BibitemShut {NoStop}%
\bibitem [{\citenamefont {Hochberg}\ \emph {et~al.}(2023)\citenamefont
  {Hochberg}, \citenamefont {Kramer}, \citenamefont {Kurinsky},\ and\
  \citenamefont {Lehmann}}]{Hochberg:2021ymx}%
  \BibitemOpen
  \bibfield  {author} {\bibinfo {author} {\bibfnamefont {Y.}~\bibnamefont
  {Hochberg}}, \bibinfo {author} {\bibfnamefont {E.~D.}\ \bibnamefont
  {Kramer}}, \bibinfo {author} {\bibfnamefont {N.}~\bibnamefont {Kurinsky}},\
  and\ \bibinfo {author} {\bibfnamefont {B.~V.}\ \bibnamefont {Lehmann}},\
  }\bibfield  {title} {\bibinfo {title} {{Directional detection of light dark
  matter in superconductors}},\ }\href
  {https://doi.org/10.1103/PhysRevD.107.076015} {\bibfield  {journal} {\bibinfo
   {journal} {Phys. Rev. D}\ }\textbf {\bibinfo {volume} {107}},\ \bibinfo
  {pages} {076015} (\bibinfo {year} {2023})},\ \Eprint
  {https://arxiv.org/abs/2109.04473} {arXiv:2109.04473 [hep-ph]} \BibitemShut
  {NoStop}%
\bibitem [{\citenamefont {Hochberg}\ \emph {et~al.}(2022)\citenamefont
  {Hochberg}, \citenamefont {Lehmann}, \citenamefont {Charaev}, \citenamefont
  {Chiles}, \citenamefont {Colangelo}, \citenamefont {Nam},\ and\ \citenamefont
  {Berggren}}]{Hochberg:2021yud}%
  \BibitemOpen
  \bibfield  {author} {\bibinfo {author} {\bibfnamefont {Y.}~\bibnamefont
  {Hochberg}}, \bibinfo {author} {\bibfnamefont {B.~V.}\ \bibnamefont
  {Lehmann}}, \bibinfo {author} {\bibfnamefont {I.}~\bibnamefont {Charaev}},
  \bibinfo {author} {\bibfnamefont {J.}~\bibnamefont {Chiles}}, \bibinfo
  {author} {\bibfnamefont {M.}~\bibnamefont {Colangelo}}, \bibinfo {author}
  {\bibfnamefont {S.~W.}\ \bibnamefont {Nam}},\ and\ \bibinfo {author}
  {\bibfnamefont {K.~K.}\ \bibnamefont {Berggren}},\ }\bibfield  {title}
  {\bibinfo {title} {{New constraints on dark matter from superconducting
  nanowires}},\ }\href {https://doi.org/10.1103/PhysRevD.106.112005} {\bibfield
   {journal} {\bibinfo  {journal} {Phys. Rev. D}\ }\textbf {\bibinfo {volume}
  {106}},\ \bibinfo {pages} {112005} (\bibinfo {year} {2022})},\ \Eprint
  {https://arxiv.org/abs/2110.01586} {arXiv:2110.01586 [hep-ph]} \BibitemShut
  {NoStop}%
\bibitem [{\citenamefont {Derenzo}\ \emph {et~al.}(2017)\citenamefont
  {Derenzo}, \citenamefont {Essig}, \citenamefont {Massari}, \citenamefont
  {Soto},\ and\ \citenamefont {Yu}}]{Derenzo:2016fse}%
  \BibitemOpen
  \bibfield  {author} {\bibinfo {author} {\bibfnamefont {S.}~\bibnamefont
  {Derenzo}}, \bibinfo {author} {\bibfnamefont {R.}~\bibnamefont {Essig}},
  \bibinfo {author} {\bibfnamefont {A.}~\bibnamefont {Massari}}, \bibinfo
  {author} {\bibfnamefont {A.}~\bibnamefont {Soto}},\ and\ \bibinfo {author}
  {\bibfnamefont {T.-T.}\ \bibnamefont {Yu}},\ }\bibfield  {title} {\bibinfo
  {title} {{Direct Detection of sub-GeV Dark Matter with Scintillating
  Targets}},\ }\href {https://doi.org/10.1103/PhysRevD.96.016026} {\bibfield
  {journal} {\bibinfo  {journal} {Phys. Rev. D}\ }\textbf {\bibinfo {volume}
  {96}},\ \bibinfo {pages} {016026} (\bibinfo {year} {2017})},\ \Eprint
  {https://arxiv.org/abs/1607.01009} {arXiv:1607.01009 [hep-ph]} \BibitemShut
  {NoStop}%
\bibitem [{\citenamefont {Hochberg}\ \emph
  {et~al.}(2017{\natexlab{a}})\citenamefont {Hochberg}, \citenamefont {Kahn},
  \citenamefont {Lisanti}, \citenamefont {Tully},\ and\ \citenamefont
  {Zurek}}]{Hochberg:2016ntt}%
  \BibitemOpen
  \bibfield  {author} {\bibinfo {author} {\bibfnamefont {Y.}~\bibnamefont
  {Hochberg}}, \bibinfo {author} {\bibfnamefont {Y.}~\bibnamefont {Kahn}},
  \bibinfo {author} {\bibfnamefont {M.}~\bibnamefont {Lisanti}}, \bibinfo
  {author} {\bibfnamefont {C.~G.}\ \bibnamefont {Tully}},\ and\ \bibinfo
  {author} {\bibfnamefont {K.~M.}\ \bibnamefont {Zurek}},\ }\bibfield  {title}
  {\bibinfo {title} {{Directional detection of dark matter with two-dimensional
  targets}},\ }\href {https://doi.org/10.1016/j.physletb.2017.06.051}
  {\bibfield  {journal} {\bibinfo  {journal} {Phys. Lett. B}\ }\textbf
  {\bibinfo {volume} {772}},\ \bibinfo {pages} {239} (\bibinfo {year}
  {2017}{\natexlab{a}})},\ \Eprint {https://arxiv.org/abs/1606.08849}
  {arXiv:1606.08849 [hep-ph]} \BibitemShut {NoStop}%
\bibitem [{\citenamefont {Hochberg}\ \emph {et~al.}(2018)\citenamefont
  {Hochberg}, \citenamefont {Kahn}, \citenamefont {Lisanti}, \citenamefont
  {Zurek}, \citenamefont {Grushin}, \citenamefont {Ilan} \emph
  {et~al.}}]{Hochberg:2017wce}%
  \BibitemOpen
  \bibfield  {author} {\bibinfo {author} {\bibfnamefont {Y.}~\bibnamefont
  {Hochberg}}, \bibinfo {author} {\bibfnamefont {Y.}~\bibnamefont {Kahn}},
  \bibinfo {author} {\bibfnamefont {M.}~\bibnamefont {Lisanti}}, \bibinfo
  {author} {\bibfnamefont {K.~M.}\ \bibnamefont {Zurek}}, \bibinfo {author}
  {\bibfnamefont {A.~G.}\ \bibnamefont {Grushin}}, \bibinfo {author}
  {\bibfnamefont {R.}~\bibnamefont {Ilan}}, \emph {et~al.},\ }\bibfield
  {title} {\bibinfo {title} {{Detection of sub-MeV Dark Matter with
  Three-Dimensional Dirac Materials}},\ }\href
  {https://doi.org/10.1103/PhysRevD.97.015004} {\bibfield  {journal} {\bibinfo
  {journal} {Phys. Rev. D}\ }\textbf {\bibinfo {volume} {97}},\ \bibinfo
  {pages} {015004} (\bibinfo {year} {2018})},\ \Eprint
  {https://arxiv.org/abs/1708.08929} {arXiv:1708.08929 [hep-ph]} \BibitemShut
  {NoStop}%
\bibitem [{\citenamefont {Cavoto}\ \emph {et~al.}(2018)\citenamefont {Cavoto},
  \citenamefont {Luchetta},\ and\ \citenamefont {Polosa}}]{Cavoto:2017otc}%
  \BibitemOpen
  \bibfield  {author} {\bibinfo {author} {\bibfnamefont {G.}~\bibnamefont
  {Cavoto}}, \bibinfo {author} {\bibfnamefont {F.}~\bibnamefont {Luchetta}},\
  and\ \bibinfo {author} {\bibfnamefont {A.~D.}\ \bibnamefont {Polosa}},\
  }\bibfield  {title} {\bibinfo {title} {{Sub-GeV Dark Matter Detection with
  Electron Recoils in Carbon Nanotubes}},\ }\href
  {https://doi.org/10.1016/j.physletb.2017.11.064} {\bibfield  {journal}
  {\bibinfo  {journal} {Phys. Lett. B}\ }\textbf {\bibinfo {volume} {776}},\
  \bibinfo {pages} {338} (\bibinfo {year} {2018})},\ \Eprint
  {https://arxiv.org/abs/1706.02487} {arXiv:1706.02487 [hep-ph]} \BibitemShut
  {NoStop}%
\bibitem [{\citenamefont {Kurinsky}\ \emph {et~al.}(2019)\citenamefont
  {Kurinsky}, \citenamefont {Yu}, \citenamefont {Hochberg},\ and\ \citenamefont
  {Cabrera}}]{Kurinsky:2019pgb}%
  \BibitemOpen
  \bibfield  {author} {\bibinfo {author} {\bibfnamefont {N.~A.}\ \bibnamefont
  {Kurinsky}}, \bibinfo {author} {\bibfnamefont {T.~C.}\ \bibnamefont {Yu}},
  \bibinfo {author} {\bibfnamefont {Y.}~\bibnamefont {Hochberg}},\ and\
  \bibinfo {author} {\bibfnamefont {B.}~\bibnamefont {Cabrera}},\ }\bibfield
  {title} {\bibinfo {title} {{Diamond Detectors for Direct Detection of Sub-GeV
  Dark Matter}},\ }\href {https://doi.org/10.1103/PhysRevD.99.123005}
  {\bibfield  {journal} {\bibinfo  {journal} {Phys. Rev. D}\ }\textbf {\bibinfo
  {volume} {99}},\ \bibinfo {pages} {123005} (\bibinfo {year} {2019})},\
  \Eprint {https://arxiv.org/abs/1901.07569} {arXiv:1901.07569 [hep-ex]}
  \BibitemShut {NoStop}%
\bibitem [{\citenamefont {Blanco}\ \emph {et~al.}(2020)\citenamefont {Blanco},
  \citenamefont {Collar}, \citenamefont {Kahn},\ and\ \citenamefont
  {Lillard}}]{Blanco:2019lrf}%
  \BibitemOpen
  \bibfield  {author} {\bibinfo {author} {\bibfnamefont {C.}~\bibnamefont
  {Blanco}}, \bibinfo {author} {\bibfnamefont {J.~I.}\ \bibnamefont {Collar}},
  \bibinfo {author} {\bibfnamefont {Y.}~\bibnamefont {Kahn}},\ and\ \bibinfo
  {author} {\bibfnamefont {B.}~\bibnamefont {Lillard}},\ }\bibfield  {title}
  {\bibinfo {title} {{Dark Matter-Electron Scattering from Aromatic Organic
  Targets}},\ }\href {https://doi.org/10.1103/PhysRevD.101.056001} {\bibfield
  {journal} {\bibinfo  {journal} {Phys. Rev. D}\ }\textbf {\bibinfo {volume}
  {101}},\ \bibinfo {pages} {056001} (\bibinfo {year} {2020})},\ \Eprint
  {https://arxiv.org/abs/1912.02822} {arXiv:1912.02822 [hep-ph]} \BibitemShut
  {NoStop}%
\bibitem [{\citenamefont {Griffin}\ \emph {et~al.}(2021)\citenamefont
  {Griffin}, \citenamefont {Hochberg}, \citenamefont {Inzani}, \citenamefont
  {Kurinsky}, \citenamefont {Lin},\ and\ \citenamefont
  {Chin}}]{Griffin:2020lgd}%
  \BibitemOpen
  \bibfield  {author} {\bibinfo {author} {\bibfnamefont {S.~M.}\ \bibnamefont
  {Griffin}}, \bibinfo {author} {\bibfnamefont {Y.}~\bibnamefont {Hochberg}},
  \bibinfo {author} {\bibfnamefont {K.}~\bibnamefont {Inzani}}, \bibinfo
  {author} {\bibfnamefont {N.}~\bibnamefont {Kurinsky}}, \bibinfo {author}
  {\bibfnamefont {T.}~\bibnamefont {Lin}},\ and\ \bibinfo {author}
  {\bibfnamefont {T.}~\bibnamefont {Chin}},\ }\bibfield  {title} {\bibinfo
  {title} {{Silicon carbide detectors for sub-GeV dark matter}},\ }\href
  {https://doi.org/10.1103/PhysRevD.103.075002} {\bibfield  {journal} {\bibinfo
   {journal} {Phys. Rev. D}\ }\textbf {\bibinfo {volume} {103}},\ \bibinfo
  {pages} {075002} (\bibinfo {year} {2021})},\ \Eprint
  {https://arxiv.org/abs/2008.08560} {arXiv:2008.08560 [hep-ph]} \BibitemShut
  {NoStop}%
\bibitem [{\citenamefont {Simchony}\ \emph {et~al.}(2024)\citenamefont
  {Simchony} \emph {et~al.}}]{Simchony:2024kcn}%
  \BibitemOpen
  \bibfield  {author} {\bibinfo {author} {\bibfnamefont {A.}~\bibnamefont
  {Simchony}} \emph {et~al.},\ }\bibfield  {title} {\bibinfo {title} {{Diamond
  and SiC Detectors for Rare Event Searches}},\ }\href
  {https://doi.org/10.1007/s10909-024-03148-4} {\bibfield  {journal} {\bibinfo
  {journal} {J. Low Temp. Phys.}\ }\textbf {\bibinfo {volume} {216}},\ \bibinfo
  {pages} {363} (\bibinfo {year} {2024})}\BibitemShut {NoStop}%
\bibitem [{\citenamefont {Essig}\ \emph {et~al.}(2022)\citenamefont {Essig}
  \emph {et~al.}}]{Essig:2022dfa}%
  \BibitemOpen
  \bibfield  {author} {\bibinfo {author} {\bibfnamefont {R.}~\bibnamefont
  {Essig}} \emph {et~al.},\ }\bibfield  {title} {\bibinfo {title}
  {{Snowmass2021 Cosmic Frontier: The landscape of low-threshold dark matter
  direct detection in the next decade}},\ }in\ \href@noop {} {\emph {\bibinfo
  {booktitle} {{Snowmass 2021}}}}\ (\bibinfo {year} {2022})\ \Eprint
  {https://arxiv.org/abs/2203.08297} {arXiv:2203.08297 [hep-ph]} \BibitemShut
  {NoStop}%
\bibitem [{\citenamefont {Das}\ \emph {et~al.}(2024{\natexlab{a}})\citenamefont
  {Das}, \citenamefont {Kurinsky},\ and\ \citenamefont {Leane}}]{Das:2022srn}%
  \BibitemOpen
  \bibfield  {author} {\bibinfo {author} {\bibfnamefont {A.}~\bibnamefont
  {Das}}, \bibinfo {author} {\bibfnamefont {N.}~\bibnamefont {Kurinsky}},\ and\
  \bibinfo {author} {\bibfnamefont {R.~K.}\ \bibnamefont {Leane}},\ }\bibfield
  {title} {\bibinfo {title} {{Dark Matter Induced Power in Quantum Devices}},\
  }\href {https://doi.org/10.1103/PhysRevLett.132.121801} {\bibfield  {journal}
  {\bibinfo  {journal} {Phys. Rev. Lett.}\ }\textbf {\bibinfo {volume} {132}},\
  \bibinfo {pages} {121801} (\bibinfo {year} {2024}{\natexlab{a}})},\ \Eprint
  {https://arxiv.org/abs/2210.09313} {arXiv:2210.09313 [hep-ph]} \BibitemShut
  {NoStop}%
\bibitem [{\citenamefont {Das}\ \emph {et~al.}(2024{\natexlab{b}})\citenamefont
  {Das}, \citenamefont {Kurinsky},\ and\ \citenamefont {Leane}}]{Das:2024jdz}%
  \BibitemOpen
  \bibfield  {author} {\bibinfo {author} {\bibfnamefont {A.}~\bibnamefont
  {Das}}, \bibinfo {author} {\bibfnamefont {N.}~\bibnamefont {Kurinsky}},\ and\
  \bibinfo {author} {\bibfnamefont {R.~K.}\ \bibnamefont {Leane}},\ }\bibfield
  {title} {\bibinfo {title} {{Transmon Qubit constraints on dark matter-nucleon
  scattering}},\ }\href {https://doi.org/10.1007/JHEP07(2024)233} {\bibfield
  {journal} {\bibinfo  {journal} {JHEP}\ }\textbf {\bibinfo {volume} {07}},\
  \bibinfo {pages} {233}},\ \Eprint {https://arxiv.org/abs/2405.00112}
  {arXiv:2405.00112 [hep-ph]} \BibitemShut {NoStop}%
\bibitem [{\citenamefont {Griffin}\ \emph {et~al.}(2025)\citenamefont
  {Griffin}, \citenamefont {Hadas}, \citenamefont {Hochberg}, \citenamefont
  {Inzani},\ and\ \citenamefont {Lehmann}}]{Griffin:2024cew}%
  \BibitemOpen
  \bibfield  {author} {\bibinfo {author} {\bibfnamefont {S.~M.}\ \bibnamefont
  {Griffin}}, \bibinfo {author} {\bibfnamefont {G.~D.}\ \bibnamefont {Hadas}},
  \bibinfo {author} {\bibfnamefont {Y.}~\bibnamefont {Hochberg}}, \bibinfo
  {author} {\bibfnamefont {K.}~\bibnamefont {Inzani}},\ and\ \bibinfo {author}
  {\bibfnamefont {B.~V.}\ \bibnamefont {Lehmann}},\ }\bibfield  {title}
  {\bibinfo {title} {{Dark Matter-Electron Detectors for Dark Matter-Nucleon
  Interactions}},\ }\href {https://doi.org/10.1103/6qqv-rl7q} {\bibfield
  {journal} {\bibinfo  {journal} {Phys. Rev. Lett.}\ }\textbf {\bibinfo
  {volume} {135}},\ \bibinfo {pages} {141803} (\bibinfo {year}
  {2025})}\BibitemShut {NoStop}%
\bibitem [{\citenamefont {Baudis}\ \emph {et~al.}(2025)\citenamefont {Baudis}
  \emph {et~al.}}]{QROCODILE:2024zmg}%
  \BibitemOpen
  \bibfield  {author} {\bibinfo {author} {\bibfnamefont {L.}~\bibnamefont
  {Baudis}} \emph {et~al.} (\bibinfo {collaboration} {QROCODILE Collaboration}),\ }\bibfield
  {title} {\bibinfo {title} {{First Sub-MeV Dark Matter Search with the
  QROCODILE Experiment Using Superconducting Nanowire Single-Photon
  Detectors}},\ }\href {https://doi.org/10.1103/4hb6-f6jl} {\bibfield
  {journal} {\bibinfo  {journal} {Phys. Rev. Lett.}\ }\textbf {\bibinfo
  {volume} {135}},\ \bibinfo {pages} {081002} (\bibinfo {year}
  {2025})}\BibitemShut {NoStop}%
\bibitem [{\citenamefont {Li}\ \emph {et~al.}(2007)\citenamefont {Li},
  \citenamefont {Hu}, \citenamefont {Qian}, \citenamefont {Hsieh},
  \citenamefont {Hasan} \emph {et~al.}}]{Li:2007}%
  \BibitemOpen
  \bibfield  {author} {\bibinfo {author} {\bibfnamefont {G.}~\bibnamefont
  {Li}}, \bibinfo {author} {\bibfnamefont {W.~Z.}\ \bibnamefont {Hu}}, \bibinfo
  {author} {\bibfnamefont {D.}~\bibnamefont {Qian}}, \bibinfo {author}
  {\bibfnamefont {D.}~\bibnamefont {Hsieh}}, \bibinfo {author} {\bibfnamefont
  {M.~Z.}\ \bibnamefont {Hasan}}, \emph {et~al.},\ }\bibfield  {title}
  {\bibinfo {title} {Semimetal-to-semimetal charge density wave transition in
  $1t$-tise$_2$},\ }\href {https://doi.org/10.1103/PhysRevLett.99.027404}
  {\bibfield  {journal} {\bibinfo  {journal} {Phys. Rev. Lett.}\ }\textbf
  {\bibinfo {volume} {99}},\ \bibinfo {pages} {027404} (\bibinfo {year}
  {2007})}\BibitemShut {NoStop}%
\bibitem [{\citenamefont {Kogar}\ \emph {et~al.}(2017)\citenamefont {Kogar},
  \citenamefont {Rak}, \citenamefont {Vig}, \citenamefont {Husain},
  \citenamefont {Flicker}, \citenamefont {Joe}, \citenamefont {Venema},
  \citenamefont {MacDougall}, \citenamefont {Chiang}, \citenamefont {Fradkin},
  \citenamefont {van Wezel},\ and\ \citenamefont {Abbamonte}}]{Kogar2017}%
  \BibitemOpen
  \bibfield  {author} {\bibinfo {author} {\bibfnamefont {A.}~\bibnamefont
  {Kogar}}, \bibinfo {author} {\bibfnamefont {M.~S.}\ \bibnamefont {Rak}},
  \bibinfo {author} {\bibfnamefont {S.}~\bibnamefont {Vig}}, \bibinfo {author}
  {\bibfnamefont {A.~A.}\ \bibnamefont {Husain}}, \bibinfo {author}
  {\bibfnamefont {F.}~\bibnamefont {Flicker}}, \bibinfo {author} {\bibfnamefont
  {Y.~I.}\ \bibnamefont {Joe}}, \bibinfo {author} {\bibfnamefont
  {L.}~\bibnamefont {Venema}}, \bibinfo {author} {\bibfnamefont {G.~J.}\
  \bibnamefont {MacDougall}}, \bibinfo {author} {\bibfnamefont {T.~C.}\
  \bibnamefont {Chiang}}, \bibinfo {author} {\bibfnamefont {E.}~\bibnamefont
  {Fradkin}}, \bibinfo {author} {\bibfnamefont {J.}~\bibnamefont {van Wezel}},\
  and\ \bibinfo {author} {\bibfnamefont {P.}~\bibnamefont {Abbamonte}},\
  }\bibfield  {title} {\bibinfo {title} {Signatures of exciton condensation in
  a transition metal dichalcogenide},\ }\href
  {https://doi.org/10.1126/science.aam6432} {\bibfield  {journal} {\bibinfo
  {journal} {Science}\ }\textbf {\bibinfo {volume} {358}},\ \bibinfo {pages}
  {1314–1317} (\bibinfo {year} {2017})}\BibitemShut {NoStop}%
\bibitem [{\citenamefont {Lin}\ \emph {et~al.}(2022)\citenamefont {Lin},
  \citenamefont {Wang}, \citenamefont {Balassis}, \citenamefont {Echeverry},
  \citenamefont {Vasenko} \emph {et~al.}}]{lin2022}%
  \BibitemOpen
  \bibfield  {author} {\bibinfo {author} {\bibfnamefont {Z.}~\bibnamefont
  {Lin}}, \bibinfo {author} {\bibfnamefont {C.}~\bibnamefont {Wang}}, \bibinfo
  {author} {\bibfnamefont {A.}~\bibnamefont {Balassis}}, \bibinfo {author}
  {\bibfnamefont {J.~P.}\ \bibnamefont {Echeverry}}, \bibinfo {author}
  {\bibfnamefont {A.~S.}\ \bibnamefont {Vasenko}}, \emph {et~al.},\ }\bibfield
  {title} {\bibinfo {title} {Dramatic plasmon response to the
  charge-density-wave gap development in $1t$-tise$_2$},\ }\href
  {https://doi.org/10.1103/PhysRevLett.129.187601} {\bibfield  {journal}
  {\bibinfo  {journal} {Phys. Rev. Lett.}\ }\textbf {\bibinfo {volume} {129}},\
  \bibinfo {pages} {187601} (\bibinfo {year} {2022})}\BibitemShut {NoStop}%
\bibitem [{\citenamefont {Husain}\ \emph {et~al.}(2023)\citenamefont {Husain},
  \citenamefont {Huang}, \citenamefont {Mitrano}, \citenamefont {Rak},
  \citenamefont {Rubeck} \emph {et~al.}}]{husain2023}%
  \BibitemOpen
  \bibfield  {author} {\bibinfo {author} {\bibfnamefont {A.~A.}\ \bibnamefont
  {Husain}}, \bibinfo {author} {\bibfnamefont {E.~W.}\ \bibnamefont {Huang}},
  \bibinfo {author} {\bibfnamefont {M.}~\bibnamefont {Mitrano}}, \bibinfo
  {author} {\bibfnamefont {M.~S.}\ \bibnamefont {Rak}}, \bibinfo {author}
  {\bibfnamefont {S.~I.}\ \bibnamefont {Rubeck}}, \emph {et~al.},\ }\bibfield
  {title} {\bibinfo {title} {Pines' demon observed as a 3d acoustic plasmon in
  sr$_2$ruo$_4$},\ }\href {https://doi.org/10.1038/s41586-023-06318-8}
  {\bibfield  {journal} {\bibinfo  {journal} {Nature}\ }\textbf {\bibinfo
  {volume} {621}},\ \bibinfo {pages} {66} (\bibinfo {year} {2023})}\BibitemShut
  {NoStop}%
\bibitem [{\citenamefont {Bhattacharya}\ \emph {et~al.}(2025)\citenamefont
  {Bhattacharya}, \citenamefont {Boyd}, \citenamefont {Reichardt},
  \citenamefont {Allard}, \citenamefont {Talebi} \emph
  {et~al.}}]{bhattacharya2025}%
  \BibitemOpen
  \bibfield  {author} {\bibinfo {author} {\bibfnamefont {S.}~\bibnamefont
  {Bhattacharya}}, \bibinfo {author} {\bibfnamefont {J.}~\bibnamefont {Boyd}},
  \bibinfo {author} {\bibfnamefont {S.}~\bibnamefont {Reichardt}}, \bibinfo
  {author} {\bibfnamefont {V.}~\bibnamefont {Allard}}, \bibinfo {author}
  {\bibfnamefont {A.~H.}\ \bibnamefont {Talebi}}, \emph {et~al.},\ }\bibfield
  {title} {\bibinfo {title} {Intervalence plasmons in boron-doped diamond},\
  }\href {https://doi.org/10.1038/s41467-024-55353-0} {\bibfield  {journal}
  {\bibinfo  {journal} {Nature Communications}\ }\textbf {\bibinfo {volume}
  {16}},\ \bibinfo {pages} {444} (\bibinfo {year} {2025})}\BibitemShut
  {NoStop}%
\bibitem [{\citenamefont {Hochberg}\ \emph {et~al.}(2021)\citenamefont
  {Hochberg}, \citenamefont {Kahn}, \citenamefont {Kurinsky}, \citenamefont
  {Lehmann}, \citenamefont {Yu},\ and\ \citenamefont
  {Berggren}}]{Hochberg:2021pkt}%
  \BibitemOpen
  \bibfield  {author} {\bibinfo {author} {\bibfnamefont {Y.}~\bibnamefont
  {Hochberg}}, \bibinfo {author} {\bibfnamefont {Y.}~\bibnamefont {Kahn}},
  \bibinfo {author} {\bibfnamefont {N.}~\bibnamefont {Kurinsky}}, \bibinfo
  {author} {\bibfnamefont {B.~V.}\ \bibnamefont {Lehmann}}, \bibinfo {author}
  {\bibfnamefont {T.~C.}\ \bibnamefont {Yu}},\ and\ \bibinfo {author}
  {\bibfnamefont {K.~K.}\ \bibnamefont {Berggren}},\ }\bibfield  {title}
  {\bibinfo {title} {{Determining Dark-Matter\textendash{}Electron Scattering
  Rates from the Dielectric Function}},\ }\href
  {https://doi.org/10.1103/PhysRevLett.127.151802} {\bibfield  {journal}
  {\bibinfo  {journal} {Phys. Rev. Lett.}\ }\textbf {\bibinfo {volume} {127}},\
  \bibinfo {pages} {151802} (\bibinfo {year} {2021})},\ \Eprint
  {https://arxiv.org/abs/2101.08263} {arXiv:2101.08263 [hep-ph]} \BibitemShut
  {NoStop}%
\bibitem [{\citenamefont {Boyd}\ \emph {et~al.}(2023)\citenamefont {Boyd},
  \citenamefont {Hochberg}, \citenamefont {Kahn}, \citenamefont {Kramer},
  \citenamefont {Kurinsky}, \citenamefont {Lehmann} \emph
  {et~al.}}]{Boyd:2022tcn}%
  \BibitemOpen
  \bibfield  {author} {\bibinfo {author} {\bibfnamefont {C.}~\bibnamefont
  {Boyd}}, \bibinfo {author} {\bibfnamefont {Y.}~\bibnamefont {Hochberg}},
  \bibinfo {author} {\bibfnamefont {Y.}~\bibnamefont {Kahn}}, \bibinfo {author}
  {\bibfnamefont {E.~D.}\ \bibnamefont {Kramer}}, \bibinfo {author}
  {\bibfnamefont {N.}~\bibnamefont {Kurinsky}}, \bibinfo {author}
  {\bibfnamefont {B.~V.}\ \bibnamefont {Lehmann}}, \emph {et~al.},\ }\bibfield
  {title} {\bibinfo {title} {{Directional detection of dark matter with
  anisotropic response functions}},\ }\href
  {https://doi.org/10.1103/PhysRevD.108.015015} {\bibfield  {journal} {\bibinfo
   {journal} {Phys. Rev. D}\ }\textbf {\bibinfo {volume} {108}},\ \bibinfo
  {pages} {015015} (\bibinfo {year} {2023})},\ \Eprint
  {https://arxiv.org/abs/2212.04505} {arXiv:2212.04505 [hep-ph]} \BibitemShut
  {NoStop}%
\bibitem [{\citenamefont {Knapen}\ \emph {et~al.}(2022)\citenamefont {Knapen},
  \citenamefont {Kozaczuk},\ and\ \citenamefont {Lin}}]{Knapen:2021bwg}%
  \BibitemOpen
  \bibfield  {author} {\bibinfo {author} {\bibfnamefont {S.}~\bibnamefont
  {Knapen}}, \bibinfo {author} {\bibfnamefont {J.}~\bibnamefont {Kozaczuk}},\
  and\ \bibinfo {author} {\bibfnamefont {T.}~\bibnamefont {Lin}},\ }\bibfield
  {title} {\bibinfo {title} {{Python package for dark matter scattering in
  dielectric targets}},\ }\href {https://doi.org/10.1103/PhysRevD.105.015014}
  {\bibfield  {journal} {\bibinfo  {journal} {Phys. Rev. D}\ }\textbf {\bibinfo
  {volume} {105}},\ \bibinfo {pages} {015014} (\bibinfo {year} {2022})},\
  \Eprint {https://arxiv.org/abs/2104.12786} {arXiv:2104.12786 [hep-ph]}
  \BibitemShut {NoStop}%
\bibitem [{\citenamefont {Knapen}\ \emph {et~al.}(2021)\citenamefont {Knapen},
  \citenamefont {Kozaczuk},\ and\ \citenamefont {Lin}}]{Knapen:2021run}%
  \BibitemOpen
  \bibfield  {author} {\bibinfo {author} {\bibfnamefont {S.}~\bibnamefont
  {Knapen}}, \bibinfo {author} {\bibfnamefont {J.}~\bibnamefont {Kozaczuk}},\
  and\ \bibinfo {author} {\bibfnamefont {T.}~\bibnamefont {Lin}},\ }\bibfield
  {title} {\bibinfo {title} {{Dark matter-electron scattering in
  dielectrics}},\ }\href {https://doi.org/10.1103/PhysRevD.104.015031}
  {\bibfield  {journal} {\bibinfo  {journal} {Phys. Rev. D}\ }\textbf {\bibinfo
  {volume} {104}},\ \bibinfo {pages} {015031} (\bibinfo {year} {2021})},\
  \Eprint {https://arxiv.org/abs/2101.08275} {arXiv:2101.08275 [hep-ph]}
  \BibitemShut {NoStop}%
\bibitem [{sup()}]{supplemental}%
  \BibitemOpen
  \href@noop {} {}\bibinfo {note} {See Supplemental Material at
  \url{http://link.aps.org/supplemental/10.1103/jdw6-3556} for additional
  details on the computational methods and material response calculations,
  which includes
  Refs.~\cite{gpaw,yan2011,dressel2002electrodynamics}}\BibitemShut {NoStop}%
\bibitem [{\citenamefont {Mortensen}\ \emph {et~al.}(2024)\citenamefont
  {Mortensen}, \citenamefont {Larsen}, \citenamefont {Kuisma}, \citenamefont
  {Ivanov}, \citenamefont {Taghizadeh}, \citenamefont {Peterson} \emph
  {et~al.}}]{gpaw}%
  \BibitemOpen
  \bibfield  {author} {\bibinfo {author} {\bibfnamefont {J.~J.}\ \bibnamefont
  {Mortensen}}, \bibinfo {author} {\bibfnamefont {A.~H.}\ \bibnamefont
  {Larsen}}, \bibinfo {author} {\bibfnamefont {M.}~\bibnamefont {Kuisma}},
  \bibinfo {author} {\bibfnamefont {A.~V.}\ \bibnamefont {Ivanov}}, \bibinfo
  {author} {\bibfnamefont {A.}~\bibnamefont {Taghizadeh}}, \bibinfo {author}
  {\bibfnamefont {A.}~\bibnamefont {Peterson}}, \emph {et~al.},\ }\bibfield
  {title} {\bibinfo {title} {{GPAW}: An open {Python} package for electronic
  structure calculations},\ }\href {https://doi.org/10.1063/5.0182685}
  {\bibfield  {journal} {\bibinfo  {journal} {The Journal of Chemical Physics}\
  }\textbf {\bibinfo {volume} {160}},\ \bibinfo {pages} {092503} (\bibinfo
  {year} {2024})}\BibitemShut {NoStop}%
\bibitem [{\citenamefont {Yan}\ \emph {et~al.}(2011)\citenamefont {Yan},
  \citenamefont {Mortensen}, \citenamefont {Jacobsen},\ and\ \citenamefont
  {Thygesen}}]{yan2011}%
  \BibitemOpen
  \bibfield  {author} {\bibinfo {author} {\bibfnamefont {J.}~\bibnamefont
  {Yan}}, \bibinfo {author} {\bibfnamefont {J.~J.}\ \bibnamefont {Mortensen}},
  \bibinfo {author} {\bibfnamefont {K.~W.}\ \bibnamefont {Jacobsen}},\ and\
  \bibinfo {author} {\bibfnamefont {K.~S.}\ \bibnamefont {Thygesen}},\
  }\bibfield  {title} {\bibinfo {title} {Linear density response function in
  the projector augmented wave method: Applications to solids, surfaces, and
  interfaces},\ }\href {https://doi.org/10.1103/PhysRevB.83.245122} {\bibfield
  {journal} {\bibinfo  {journal} {Phys. Rev. B}\ }\textbf {\bibinfo {volume}
  {83}},\ \bibinfo {pages} {245122} (\bibinfo {year} {2011})}\BibitemShut
  {NoStop}%
\bibitem [{\citenamefont {Dressel}\ and\ \citenamefont
  {Gr{\"u}ner}(2002)}]{dressel2002electrodynamics}%
  \BibitemOpen
  \bibfield  {author} {\bibinfo {author} {\bibfnamefont {M.}~\bibnamefont
  {Dressel}}\ and\ \bibinfo {author} {\bibfnamefont {G.}~\bibnamefont
  {Gr{\"u}ner}},\ }\href {https://books.google.com/books?id=Wv-wrtRdWT0C}
  {\emph {\bibinfo {title} {Electrodynamics of Solids: Optical Properties of
  Electrons in Matter}}}\ (\bibinfo  {publisher} {Cambridge University Press},\
  \bibinfo {address} {Cambridge, England},\ \bibinfo {year} {2002})\BibitemShut
  {NoStop}%
\bibitem [{\citenamefont {Yin}\ \emph {et~al.}(2024)\citenamefont {Yin},
  \citenamefont {Tang}, \citenamefont {Berlijn},\ and\ \citenamefont
  {Ruzsinszky}}]{yin2024}%
  \BibitemOpen
  \bibfield  {author} {\bibinfo {author} {\bibfnamefont {L.}~\bibnamefont
  {Yin}}, \bibinfo {author} {\bibfnamefont {H.}~\bibnamefont {Tang}}, \bibinfo
  {author} {\bibfnamefont {T.}~\bibnamefont {Berlijn}},\ and\ \bibinfo {author}
  {\bibfnamefont {A.}~\bibnamefont {Ruzsinszky}},\ }\bibfield  {title}
  {\bibinfo {title} {Efficient simulations of charge density waves in the
  transition metal dichalcogenide {TiSe2}},\ }\href
  {https://doi.org/10.1038/s41524-024-01396-2} {\bibfield  {journal} {\bibinfo
  {journal} {npj Computational Materials}\ }\textbf {\bibinfo {volume} {10}},\
  \bibinfo {pages} {207} (\bibinfo {year} {2024})}\BibitemShut {NoStop}%
\bibitem [{\citenamefont {Di~Salvo}\ \emph {et~al.}(1976)\citenamefont
  {Di~Salvo}, \citenamefont {Moncton},\ and\ \citenamefont
  {Waszczak}}]{disalvo1976}%
  \BibitemOpen
  \bibfield  {author} {\bibinfo {author} {\bibfnamefont {F.~J.}\ \bibnamefont
  {Di~Salvo}}, \bibinfo {author} {\bibfnamefont {D.~E.}\ \bibnamefont
  {Moncton}},\ and\ \bibinfo {author} {\bibfnamefont {J.~V.}\ \bibnamefont
  {Waszczak}},\ }\bibfield  {title} {\bibinfo {title} {Electronic properties
  and superlattice formation in the semimetal tise$_2$},\ }\href
  {https://doi.org/10.1103/PhysRevB.14.4321} {\bibfield  {journal} {\bibinfo
  {journal} {Phys. Rev. B}\ }\textbf {\bibinfo {volume} {14}},\ \bibinfo
  {pages} {4321} (\bibinfo {year} {1976})}\BibitemShut {NoStop}%
\bibitem [{\citenamefont {Knowles}\ \emph {et~al.}(2020)\citenamefont
  {Knowles}, \citenamefont {Yang}, \citenamefont {Muramatsu}, \citenamefont
  {Moulding}, \citenamefont {Buhot} \emph {et~al.}}]{knowles2020}%
  \BibitemOpen
  \bibfield  {author} {\bibinfo {author} {\bibfnamefont {P.}~\bibnamefont
  {Knowles}}, \bibinfo {author} {\bibfnamefont {B.}~\bibnamefont {Yang}},
  \bibinfo {author} {\bibfnamefont {T.}~\bibnamefont {Muramatsu}}, \bibinfo
  {author} {\bibfnamefont {O.}~\bibnamefont {Moulding}}, \bibinfo {author}
  {\bibfnamefont {J.}~\bibnamefont {Buhot}}, \emph {et~al.},\ }\bibfield
  {title} {\bibinfo {title} {Fermi surface reconstruction and electron dynamics
  at the charge-density-wave transition in tise$_2$},\ }\href
  {https://doi.org/10.1103/PhysRevLett.124.167602} {\bibfield  {journal}
  {\bibinfo  {journal} {Phys. Rev. Lett.}\ }\textbf {\bibinfo {volume} {124}},\
  \bibinfo {pages} {167602} (\bibinfo {year} {2020})}\BibitemShut {NoStop}%
\bibitem [{\citenamefont {Watson}\ \emph {et~al.}(2019)\citenamefont {Watson},
  \citenamefont {Clark}, \citenamefont {Mazzola}, \citenamefont
  {Markovi\ifmmode~\acute{c}\else \'{c}\fi{}}, \citenamefont {Sunko},
  \citenamefont {Kim}, \citenamefont {Rossnagel},\ and\ \citenamefont
  {King}}]{watson2019}%
  \BibitemOpen
  \bibfield  {author} {\bibinfo {author} {\bibfnamefont {M.~D.}\ \bibnamefont
  {Watson}}, \bibinfo {author} {\bibfnamefont {O.~J.}\ \bibnamefont {Clark}},
  \bibinfo {author} {\bibfnamefont {F.}~\bibnamefont {Mazzola}}, \bibinfo
  {author} {\bibfnamefont {I.}~\bibnamefont {Markovi\ifmmode~\acute{c}\else
  \'{c}\fi{}}}, \bibinfo {author} {\bibfnamefont {V.}~\bibnamefont {Sunko}},
  \bibinfo {author} {\bibfnamefont {T.~K.}\ \bibnamefont {Kim}}, \bibinfo
  {author} {\bibfnamefont {K.}~\bibnamefont {Rossnagel}},\ and\ \bibinfo
  {author} {\bibfnamefont {P.~D.~C.}\ \bibnamefont {King}},\ }\bibfield
  {title} {\bibinfo {title} {Orbital- and ${k}_{z}$-selective hybridization of
  se $4p$ and ti $3d$ states in the charge density wave phase of
  ${\mathrm{tise}}_{2}$},\ }\href
  {https://doi.org/10.1103/PhysRevLett.122.076404} {\bibfield  {journal}
  {\bibinfo  {journal} {Phys. Rev. Lett.}\ }\textbf {\bibinfo {volume} {122}},\
  \bibinfo {pages} {076404} (\bibinfo {year} {2019})}\BibitemShut {NoStop}%
\bibitem [{\citenamefont {Maeno}\ \emph {et~al.}(2024)\citenamefont {Maeno},
  \citenamefont {Ikeda},\ and\ \citenamefont {Mattoni}}]{maeno2024}%
  \BibitemOpen
  \bibfield  {author} {\bibinfo {author} {\bibfnamefont {Y.}~\bibnamefont
  {Maeno}}, \bibinfo {author} {\bibfnamefont {A.}~\bibnamefont {Ikeda}},\ and\
  \bibinfo {author} {\bibfnamefont {G.}~\bibnamefont {Mattoni}},\ }\bibfield
  {title} {\bibinfo {title} {Thirty years of puzzling superconductivity in
  sr$_2$ruo$_4$},\ }\href {https://doi.org/10.1038/s41567-024-02656-0}
  {\bibfield  {journal} {\bibinfo  {journal} {Nature Physics}\ }\textbf
  {\bibinfo {volume} {20}},\ \bibinfo {pages} {1712} (\bibinfo {year}
  {2024})}\BibitemShut {NoStop}%
\bibitem [{\citenamefont {Schultz}\ \emph {et~al.}(2025)\citenamefont
  {Schultz}, \citenamefont {Lubk}, \citenamefont {Jerzembeck}, \citenamefont
  {Kikugawa}, \citenamefont {Knupfer}, \citenamefont {Wolf}, \citenamefont
  {B{\"u}chner},\ and\ \citenamefont {Fink}}]{schultz2024}%
  \BibitemOpen
  \bibfield  {author} {\bibinfo {author} {\bibfnamefont {J.}~\bibnamefont
  {Schultz}}, \bibinfo {author} {\bibfnamefont {A.}~\bibnamefont {Lubk}},
  \bibinfo {author} {\bibfnamefont {F.}~\bibnamefont {Jerzembeck}}, \bibinfo
  {author} {\bibfnamefont {N.}~\bibnamefont {Kikugawa}}, \bibinfo {author}
  {\bibfnamefont {M.}~\bibnamefont {Knupfer}}, \bibinfo {author} {\bibfnamefont
  {D.}~\bibnamefont {Wolf}}, \bibinfo {author} {\bibfnamefont {B.}~\bibnamefont
  {B{\"u}chner}},\ and\ \bibinfo {author} {\bibfnamefont {J.}~\bibnamefont
  {Fink}},\ }\bibfield  {title} {\bibinfo {title} {Optical and acoustic
  plasmons in the layered material {Sr$_2$RuO$_4$}},\ }\href
  {https://doi.org/10.1038/s41467-025-58978-x} {\bibfield  {journal} {\bibinfo
  {journal} {Nature Communications}\ }\textbf {\bibinfo {volume} {16}},\
  \bibinfo {pages} {4287} (\bibinfo {year} {2025})}\BibitemShut {NoStop}%
\bibitem [{\citenamefont {Ekimov}\ \emph {et~al.}(2004)\citenamefont {Ekimov},
  \citenamefont {Sidorov}, \citenamefont {Bauer}, \citenamefont {Mel'nik},
  \citenamefont {Curro} \emph {et~al.}}]{ekimov2004}%
  \BibitemOpen
  \bibfield  {author} {\bibinfo {author} {\bibfnamefont {E.~A.}\ \bibnamefont
  {Ekimov}}, \bibinfo {author} {\bibfnamefont {V.~A.}\ \bibnamefont {Sidorov}},
  \bibinfo {author} {\bibfnamefont {E.~D.}\ \bibnamefont {Bauer}}, \bibinfo
  {author} {\bibfnamefont {N.~N.}\ \bibnamefont {Mel'nik}}, \bibinfo {author}
  {\bibfnamefont {N.~J.}\ \bibnamefont {Curro}}, \emph {et~al.},\ }\bibfield
  {title} {\bibinfo {title} {Superconductivity in diamond},\ }\href
  {https://doi.org/10.1038/nature02449} {\bibfield  {journal} {\bibinfo
  {journal} {Nature}\ }\textbf {\bibinfo {volume} {428}},\ \bibinfo {pages}
  {542} (\bibinfo {year} {2004})}\BibitemShut {NoStop}%
\bibitem [{\citenamefont {Pines}(1956)}]{pines1956}%
  \BibitemOpen
  \bibfield  {author} {\bibinfo {author} {\bibfnamefont {D.}~\bibnamefont
  {Pines}},\ }\bibfield  {title} {\bibinfo {title} {Collective energy losses in
  solids},\ }\href {https://doi.org/10.1103/RevModPhys.28.184} {\bibfield
  {journal} {\bibinfo  {journal} {Rev. Mod. Phys.}\ }\textbf {\bibinfo {volume}
  {28}},\ \bibinfo {pages} {184} (\bibinfo {year} {1956})}\BibitemShut
  {NoStop}%
\bibitem [{\citenamefont {Du}\ \emph {et~al.}(2024)\citenamefont {Du},
  \citenamefont {Ega{\~n}a-Ugrinovic}, \citenamefont {Essig},\ and\
  \citenamefont {Sholapurkar}}]{Du:2022dxf}%
  \BibitemOpen
  \bibfield  {author} {\bibinfo {author} {\bibfnamefont {P.}~\bibnamefont
  {Du}}, \bibinfo {author} {\bibfnamefont {D.}~\bibnamefont
  {Ega{\~n}a-Ugrinovic}}, \bibinfo {author} {\bibfnamefont {R.}~\bibnamefont
  {Essig}},\ and\ \bibinfo {author} {\bibfnamefont {M.}~\bibnamefont
  {Sholapurkar}},\ }\bibfield  {title} {\bibinfo {title} {{Doped semiconductor
  devices for sub-MeV dark matter detection}},\ }\href
  {https://doi.org/10.1103/PhysRevD.109.055009} {\bibfield  {journal} {\bibinfo
   {journal} {Phys. Rev. D}\ }\textbf {\bibinfo {volume} {109}},\ \bibinfo
  {pages} {055009} (\bibinfo {year} {2024})},\ \Eprint
  {https://arxiv.org/abs/2212.04504} {arXiv:2212.04504 [hep-ph]} \BibitemShut
  {NoStop}%
\bibitem [{\citenamefont {Chen}\ \emph {et~al.}(2015)\citenamefont {Chen},
  \citenamefont {Reich}, \citenamefont {Kramer}, \citenamefont {Fu},
  \citenamefont {Kortshagen},\ and\ \citenamefont {Shklovskii}}]{chen2015}%
  \BibitemOpen
  \bibfield  {author} {\bibinfo {author} {\bibfnamefont {T.}~\bibnamefont
  {Chen}}, \bibinfo {author} {\bibfnamefont {K.~V.}\ \bibnamefont {Reich}},
  \bibinfo {author} {\bibfnamefont {N.~J.}\ \bibnamefont {Kramer}}, \bibinfo
  {author} {\bibfnamefont {H.}~\bibnamefont {Fu}}, \bibinfo {author}
  {\bibfnamefont {U.~R.}\ \bibnamefont {Kortshagen}},\ and\ \bibinfo {author}
  {\bibfnamefont {B.~I.}\ \bibnamefont {Shklovskii}},\ }\bibfield  {title}
  {\bibinfo {title} {Metal–insulator transition in films of doped
  semiconductor nanocrystals},\ }\href {https://doi.org/10.1038/nmat4486}
  {\bibfield  {journal} {\bibinfo  {journal} {Nature Materials}\ }\textbf
  {\bibinfo {volume} {15}},\ \bibinfo {pages} {299} (\bibinfo {year}
  {2015})}\BibitemShut {NoStop}%
\bibitem [{\citenamefont {Lewin}\ and\ \citenamefont
  {Smith}(1996)}]{Lewin:1995rx}%
  \BibitemOpen
  \bibfield  {author} {\bibinfo {author} {\bibfnamefont {J.~D.}\ \bibnamefont
  {Lewin}}\ and\ \bibinfo {author} {\bibfnamefont {P.~F.}\ \bibnamefont
  {Smith}},\ }\bibfield  {title} {\bibinfo {title} {{Review of mathematics,
  numerical factors, and corrections for dark matter experiments based on
  elastic nuclear recoil}},\ }\href
  {https://doi.org/10.1016/S0927-6505(96)00047-3} {\bibfield  {journal}
  {\bibinfo  {journal} {Astropart. Phys.}\ }\textbf {\bibinfo {volume} {6}},\
  \bibinfo {pages} {87} (\bibinfo {year} {1996})}\BibitemShut {NoStop}%
\bibitem [{\citenamefont {Hochberg}\ \emph
  {et~al.}(2017{\natexlab{b}})\citenamefont {Hochberg}, \citenamefont {Lin},\
  and\ \citenamefont {Zurek}}]{Hochberg:2016sqx}%
  \BibitemOpen
  \bibfield  {author} {\bibinfo {author} {\bibfnamefont {Y.}~\bibnamefont
  {Hochberg}}, \bibinfo {author} {\bibfnamefont {T.}~\bibnamefont {Lin}},\ and\
  \bibinfo {author} {\bibfnamefont {K.~M.}\ \bibnamefont {Zurek}},\ }\bibfield
  {title} {\bibinfo {title} {{Absorption of light dark matter in
  semiconductors}},\ }\href {https://doi.org/10.1103/PhysRevD.95.023013}
  {\bibfield  {journal} {\bibinfo  {journal} {Phys. Rev. D}\ }\textbf {\bibinfo
  {volume} {95}},\ \bibinfo {pages} {023013} (\bibinfo {year}
  {2017}{\natexlab{b}})},\ \Eprint {https://arxiv.org/abs/1608.01994}
  {arXiv:1608.01994 [hep-ph]} \BibitemShut {NoStop}%
\bibitem [{\citenamefont {Feldman}\ and\ \citenamefont
  {Cousins}(1998)}]{Feldman:1997qc}%
  \BibitemOpen
  \bibfield  {author} {\bibinfo {author} {\bibfnamefont {G.~J.}\ \bibnamefont
  {Feldman}}\ and\ \bibinfo {author} {\bibfnamefont {R.~D.}\ \bibnamefont
  {Cousins}},\ }\bibfield  {title} {\bibinfo {title} {{A Unified approach to
  the classical statistical analysis of small signals}},\ }\href
  {https://doi.org/10.1103/PhysRevD.57.3873} {\bibfield  {journal} {\bibinfo
  {journal} {Phys. Rev. D}\ }\textbf {\bibinfo {volume} {57}},\ \bibinfo
  {pages} {3873} (\bibinfo {year} {1998})},\ \Eprint
  {https://arxiv.org/abs/physics/9711021} {arXiv:physics/9711021} \BibitemShut
  {NoStop}%
\bibitem [{\citenamefont {Griffin}\ \emph {et~al.}(2026)\citenamefont
  {Griffin}, \citenamefont {Hochberg}, \citenamefont {Lehmann}, \citenamefont
  {Ovadia}, \citenamefont {Persson}, \citenamefont {Suter}, \citenamefont
  {Yang},\ and\ \citenamefont {Zhao}}]{Griffin:2025wew}%
  \BibitemOpen
  \bibfield  {author} {\bibinfo {author} {\bibfnamefont {S.~M.}\ \bibnamefont
  {Griffin}}, \bibinfo {author} {\bibfnamefont {Y.}~\bibnamefont {Hochberg}},
  \bibinfo {author} {\bibfnamefont {B.~V.}\ \bibnamefont {Lehmann}}, \bibinfo
  {author} {\bibfnamefont {R.}~\bibnamefont {Ovadia}}, \bibinfo {author}
  {\bibfnamefont {K.~A.}\ \bibnamefont {Persson}}, \bibinfo {author}
  {\bibfnamefont {B.~A.}\ \bibnamefont {Suter}}, \bibinfo {author}
  {\bibfnamefont {R.~X.}\ \bibnamefont {Yang}},\ and\ \bibinfo {author}
  {\bibfnamefont {W.}~\bibnamefont {Zhao}},\ }\bibfield  {title} {\bibinfo
  {title} {{First High-Throughput Evaluation of Dark Matter Detector
  Materials}},\ }\href {https://doi.org/10.1103/bb4g-7939} {\bibfield
  {journal} {\bibinfo  {journal} {Phys. Rev. Lett.}\ }\textbf {\bibinfo
  {volume} {136}},\ \bibinfo {pages} {191801} (\bibinfo {year}
  {2026})}\BibitemShut {NoStop}%
\bibitem [{\citenamefont {Aguilar-Arevalo}\ \emph {et~al.}(2019)\citenamefont
  {Aguilar-Arevalo} \emph {et~al.}}]{DAMIC:2019dcn}%
  \BibitemOpen
  \bibfield  {author} {\bibinfo {author} {\bibfnamefont {A.}~\bibnamefont
  {Aguilar-Arevalo}} \emph {et~al.} (\bibinfo {collaboration} {DAMIC Collaboration}),\
  }\bibfield  {title} {\bibinfo {title} {{Constraints on Light Dark Matter
  Particles Interacting with Electrons from DAMIC at SNOLAB}},\ }\href
  {https://doi.org/10.1103/PhysRevLett.123.181802} {\bibfield  {journal}
  {\bibinfo  {journal} {Phys. Rev. Lett.}\ }\textbf {\bibinfo {volume} {123}},\
  \bibinfo {pages} {181802} (\bibinfo {year} {2019})},\ \Eprint
  {https://arxiv.org/abs/1907.12628} {arXiv:1907.12628 [astro-ph.CO]}
  \BibitemShut {NoStop}%
\bibitem [{\citenamefont {Arnquist}\ \emph {et~al.}(2023)\citenamefont
  {Arnquist} \emph {et~al.}}]{DAMIC-M:2023gxo}%
  \BibitemOpen
  \bibfield  {author} {\bibinfo {author} {\bibfnamefont {I.}~\bibnamefont
  {Arnquist}} \emph {et~al.} (\bibinfo {collaboration} {DAMIC-M Collaboration}),\ }\bibfield
  {title} {\bibinfo {title} {{First Constraints from DAMIC-M on Sub-GeV
  Dark-Matter Particles Interacting with Electrons}},\ }\href
  {https://doi.org/10.1103/PhysRevLett.130.171003} {\bibfield  {journal}
  {\bibinfo  {journal} {Phys. Rev. Lett.}\ }\textbf {\bibinfo {volume} {130}},\
  \bibinfo {pages} {171003} (\bibinfo {year} {2023})},\ \Eprint
  {https://arxiv.org/abs/2302.02372} {arXiv:2302.02372 [hep-ex]} \BibitemShut
  {NoStop}%
\bibitem [{\citenamefont {Adari}\ \emph {et~al.}(2025)\citenamefont {Adari}
  \emph {et~al.}}]{SENSEI:2023zdf}%
  \BibitemOpen
  \bibfield  {author} {\bibinfo {author} {\bibfnamefont {P.}~\bibnamefont
  {Adari}} \emph {et~al.} (\bibinfo {collaboration} {SENSEI Collaboration}),\ }\bibfield
  {title} {\bibinfo {title} {{First Direct-Detection Results on Sub-GeV Dark
  Matter Using the SENSEI Detector at SNOLAB}},\ }\href
  {https://doi.org/10.1103/PhysRevLett.134.011804} {\bibfield  {journal}
  {\bibinfo  {journal} {Phys. Rev. Lett.}\ }\textbf {\bibinfo {volume} {134}},\
  \bibinfo {pages} {011804} (\bibinfo {year} {2025})},\ \Eprint
  {https://arxiv.org/abs/2312.13342} {arXiv:2312.13342 [astro-ph.CO]}
  \BibitemShut {NoStop}%
\bibitem [{Mat({\natexlab{a}})}]{MaterialsProject:mp-19338}%
  \BibitemOpen
  \bibinfo {note}
  {\url{https://next-gen.materialsproject.org/materials/mp-19338}}\BibitemShut
  {NoStop}%
\bibitem [{Mat({\natexlab{b}})}]{MaterialsProject:mp-569304}%
  \BibitemOpen
  \bibinfo {note}
  {\url{https://next-gen.materialsproject.org/materials/mp-569304}}\BibitemShut
  {NoStop}%
\bibitem [{\citenamefont {Gjerding}\ \emph {et~al.}(2017)\citenamefont
  {Gjerding}, \citenamefont {Pandey},\ and\ \citenamefont
  {Thygesen}}]{gjerding2017}%
  \BibitemOpen
  \bibfield  {author} {\bibinfo {author} {\bibfnamefont {M.~N.}\ \bibnamefont
  {Gjerding}}, \bibinfo {author} {\bibfnamefont {M.}~\bibnamefont {Pandey}},\
  and\ \bibinfo {author} {\bibfnamefont {K.~S.}\ \bibnamefont {Thygesen}},\
  }\bibfield  {title} {\bibinfo {title} {Band structure engineered layered
  metals for low-loss plasmonics},\ }\href
  {https://doi.org/10.1038/ncomms15133} {\bibfield  {journal} {\bibinfo
  {journal} {Nature Communications}\ }\textbf {\bibinfo {volume} {8}},\
  \bibinfo {pages} {15133} (\bibinfo {year} {2017})}\BibitemShut {NoStop}%
\bibitem [{\citenamefont {Nag}\ \emph {et~al.}(2020)\citenamefont {Nag},
  \citenamefont {Zhu}, \citenamefont {Bejas}, \citenamefont {Li}, \citenamefont
  {Robarts}, \citenamefont {Yamase} \emph {et~al.}}]{nag2020}%
  \BibitemOpen
  \bibfield  {author} {\bibinfo {author} {\bibfnamefont {A.}~\bibnamefont
  {Nag}}, \bibinfo {author} {\bibfnamefont {M.}~\bibnamefont {Zhu}}, \bibinfo
  {author} {\bibfnamefont {M.}~\bibnamefont {Bejas}}, \bibinfo {author}
  {\bibfnamefont {J.}~\bibnamefont {Li}}, \bibinfo {author} {\bibfnamefont
  {H.~C.}\ \bibnamefont {Robarts}}, \bibinfo {author} {\bibfnamefont
  {H.}~\bibnamefont {Yamase}}, \emph {et~al.},\ }\bibfield  {title} {\bibinfo
  {title} {Detection of acoustic plasmons in hole-doped lanthanum and bismuth
  cuprate superconductors using resonant inelastic x-ray scattering},\ }\href
  {https://doi.org/10.1103/PhysRevLett.125.257002} {\bibfield  {journal}
  {\bibinfo  {journal} {Phys. Rev. Lett.}\ }\textbf {\bibinfo {volume} {125}},\
  \bibinfo {pages} {257002} (\bibinfo {year} {2020})}\BibitemShut {NoStop}%
\bibitem [{\citenamefont {Hotje}\ \emph {et~al.}(2005)\citenamefont {Hotje},
  \citenamefont {Wartchow},\ and\ \citenamefont {Binnewies}}]{Hotje:2005xxxx}%
  \BibitemOpen
  \bibfield  {author} {\bibinfo {author} {\bibfnamefont {U.}~\bibnamefont
  {Hotje}}, \bibinfo {author} {\bibfnamefont {R.}~\bibnamefont {Wartchow}},\
  and\ \bibinfo {author} {\bibfnamefont {M.}~\bibnamefont {Binnewies}},\
  }\bibfield  {title} {\bibinfo {title} {Chemischer transport fester
  l\"{o}sungen. 15 [1]. der chemische transport von mischphasen in systemen
  tis2/tise2, tis2/tas2, tise2/tase2 und tas2/tase2},\ }\href
  {https://doi.org/10.1002/zaac.200400398} {\bibfield  {journal} {\bibinfo
  {journal} {Zeitschrift f\"{u}r anorganische und allgemeine Chemie}\ }\textbf
  {\bibinfo {volume} {631}},\ \bibinfo {pages} {403–410} (\bibinfo {year}
  {2005})}\BibitemShut {NoStop}%
\bibitem [{\citenamefont {Ubaldini}\ and\ \citenamefont
  {Giannini}(2014)}]{Ubaldini:2014xxxx}%
  \BibitemOpen
  \bibfield  {author} {\bibinfo {author} {\bibfnamefont {A.}~\bibnamefont
  {Ubaldini}}\ and\ \bibinfo {author} {\bibfnamefont {E.}~\bibnamefont
  {Giannini}},\ }\bibfield  {title} {\bibinfo {title} {Improved chemical vapor
  transport growth of transition metal dichalcogenides},\ }\href
  {https://doi.org/10.1016/j.jcrysgro.2013.12.070} {\bibfield  {journal}
  {\bibinfo  {journal} {Journal of Crystal Growth}\ }\textbf {\bibinfo {volume}
  {401}},\ \bibinfo {pages} {878–882} (\bibinfo {year} {2014})}\BibitemShut
  {NoStop}%
\bibitem [{\citenamefont {Wang}\ \emph {et~al.}(2016)\citenamefont {Wang},
  \citenamefont {Zheng}, \citenamefont {Xu}, \citenamefont {Sun}, \citenamefont
  {Hu}, \citenamefont {Lu}, \citenamefont {Liu}, \citenamefont {Zheng},
  \citenamefont {Tao},\ and\ \citenamefont {Jiao}}]{Wang:2016xxxx}%
  \BibitemOpen
  \bibfield  {author} {\bibinfo {author} {\bibfnamefont {J.}~\bibnamefont
  {Wang}}, \bibinfo {author} {\bibfnamefont {H.}~\bibnamefont {Zheng}},
  \bibinfo {author} {\bibfnamefont {G.}~\bibnamefont {Xu}}, \bibinfo {author}
  {\bibfnamefont {L.}~\bibnamefont {Sun}}, \bibinfo {author} {\bibfnamefont
  {D.}~\bibnamefont {Hu}}, \bibinfo {author} {\bibfnamefont {Z.}~\bibnamefont
  {Lu}}, \bibinfo {author} {\bibfnamefont {L.}~\bibnamefont {Liu}}, \bibinfo
  {author} {\bibfnamefont {J.}~\bibnamefont {Zheng}}, \bibinfo {author}
  {\bibfnamefont {C.}~\bibnamefont {Tao}},\ and\ \bibinfo {author}
  {\bibfnamefont {L.}~\bibnamefont {Jiao}},\ }\bibfield  {title} {\bibinfo
  {title} {Controlled synthesis of two-dimensional 1t-tise2 with charge density
  wave transition by chemical vapor transport},\ }\href
  {https://doi.org/10.1021/jacs.6b10414} {\bibfield  {journal} {\bibinfo
  {journal} {Journal of the American Chemical Society}\ }\textbf {\bibinfo
  {volume} {138}},\ \bibinfo {pages} {16216–16219} (\bibinfo {year}
  {2016})}\BibitemShut {NoStop}%
\bibitem [{\citenamefont {Yang}\ \emph {et~al.}(2025)\citenamefont {Yang},
  \citenamefont {Wu}, \citenamefont {Yu}, \citenamefont {Hou}, \citenamefont
  {Zhang}, \citenamefont {Wang}, \citenamefont {Zou}, \citenamefont {Nie},
  \citenamefont {Leng}, \citenamefont {Zhou},\ and\ \citenamefont
  {Jiang}}]{Yang:2025xxxx}%
  \BibitemOpen
  \bibfield  {author} {\bibinfo {author} {\bibfnamefont {Z.}~\bibnamefont
  {Yang}}, \bibinfo {author} {\bibfnamefont {J.}~\bibnamefont {Wu}}, \bibinfo
  {author} {\bibfnamefont {Q.}~\bibnamefont {Yu}}, \bibinfo {author}
  {\bibfnamefont {X.}~\bibnamefont {Hou}}, \bibinfo {author} {\bibfnamefont
  {Z.}~\bibnamefont {Zhang}}, \bibinfo {author} {\bibfnamefont
  {X.}~\bibnamefont {Wang}}, \bibinfo {author} {\bibfnamefont {J.}~\bibnamefont
  {Zou}}, \bibinfo {author} {\bibfnamefont {Z.}~\bibnamefont {Nie}}, \bibinfo
  {author} {\bibfnamefont {J.}~\bibnamefont {Leng}}, \bibinfo {author}
  {\bibfnamefont {P.}~\bibnamefont {Zhou}},\ and\ \bibinfo {author}
  {\bibfnamefont {Z.}~\bibnamefont {Jiang}},\ }\bibfield  {title} {\bibinfo
  {title} {Large-area high-yield 1t-tise2 saturable absorber for
  wavelength-tunable ultrafast fiber lasers},\ }\href
  {https://doi.org/10.1016/j.mtphys.2025.101770} {\bibfield  {journal}
  {\bibinfo  {journal} {Materials Today Physics}\ }\textbf {\bibinfo {volume}
  {56}},\ \bibinfo {pages} {101770} (\bibinfo {year} {2025})}\BibitemShut
  {NoStop}%
\bibitem [{\citenamefont {Banerjee}\ \emph {et~al.}(2005)\citenamefont
  {Banerjee}, \citenamefont {Dhillon}, \citenamefont {Ghosh},\ and\
  \citenamefont {Sra}}]{JMLR:v6:banerjee05a}%
  \BibitemOpen
  \bibfield  {author} {\bibinfo {author} {\bibfnamefont {A.}~\bibnamefont
  {Banerjee}}, \bibinfo {author} {\bibfnamefont {I.~S.}\ \bibnamefont
  {Dhillon}}, \bibinfo {author} {\bibfnamefont {J.}~\bibnamefont {Ghosh}},\
  and\ \bibinfo {author} {\bibfnamefont {S.}~\bibnamefont {Sra}},\ }\bibfield
  {title} {\bibinfo {title} {Clustering on the unit hypersphere using von
  Mises--Fisher distributions},\ }\href
  {https://www.jmlr.org/papers/v6/banerjee05a.html} {\bibfield  {journal}
  {\bibinfo  {journal} {Journal of Machine Learning Research}\ }\textbf
  {\bibinfo {volume} {6}},\ \bibinfo {pages} {1345} (\bibinfo {year}
  {2005})}\BibitemShut {NoStop}%
\end{thebibliography}%


\begin{thebibliography}{10}%
\makeatletter
\providecommand \@ifxundefined [1]{%
 \@ifx{#1\undefined}
}%
\providecommand \@ifnum [1]{%
 \ifnum #1\expandafter \@firstoftwo
 \else \expandafter \@secondoftwo
 \fi
}%
\providecommand \@ifx [1]{%
 \ifx #1\expandafter \@firstoftwo
 \else \expandafter \@secondoftwo
 \fi
}%
\providecommand \natexlab [1]{#1}%
\providecommand \enquote  [1]{``#1''}%
\providecommand \bibnamefont  [1]{#1}%
\providecommand \bibfnamefont [1]{#1}%
\providecommand \citenamefont [1]{#1}%
\providecommand \href@noop [0]{\@secondoftwo}%
\providecommand \href [0]{\begingroup \@sanitize@url \@href}%
\providecommand \@href[1]{\@@startlink{#1}\@@href}%
\providecommand \@@href[1]{\endgroup#1\@@endlink}%
\providecommand \@sanitize@url [0]{\catcode `\\12\catcode `\$12\catcode
  `\&12\catcode `\#12\catcode `\^12\catcode `\_12\catcode `\%12\relax}%
\providecommand \@@startlink[1]{}%
\providecommand \@@endlink[0]{}%
\providecommand \url  [0]{\begingroup\@sanitize@url \@url }%
\providecommand \@url [1]{\endgroup\@href {#1}{\urlprefix }}%
\providecommand \urlprefix  [0]{URL }%
\providecommand \Eprint [0]{\href }%
\providecommand \doibase [0]{https://doi.org/}%
\providecommand \selectlanguage [0]{\@gobble}%
\providecommand \bibinfo  [0]{\@secondoftwo}%
\providecommand \bibfield  [0]{\@secondoftwo}%
\providecommand \translation [1]{[#1]}%
\providecommand \BibitemOpen [0]{}%
\providecommand \bibitemStop [0]{}%
\providecommand \bibitemNoStop [0]{.\EOS\space}%
\providecommand \EOS [0]{\spacefactor3000\relax}%
\providecommand \BibitemShut  [1]{\csname bibitem#1\endcsname}%
\let\auto@bib@innerbib\@empty
%</preamble>
\bibitem[23]{sm:Li:2007}%
  \BibitemOpen
  \bibfield  {author} {\bibinfo {author} {\bibfnamefont {G.}~\bibnamefont
  {Li}}, \bibinfo {author} {\bibfnamefont {W.~Z.}\ \bibnamefont {Hu}}, \bibinfo
  {author} {\bibfnamefont {D.}~\bibnamefont {Qian}}, \bibinfo {author}
  {\bibfnamefont {D.}~\bibnamefont {Hsieh}}, \bibinfo {author} {\bibfnamefont
  {M.~Z.}\ \bibnamefont {Hasan}}, \emph {et~al.},\ }\bibfield  {title}
  {\bibinfo {title} {Semimetal-to-semimetal charge density wave transition in
  $1t$-tise$_2$},\ }\href {https://doi.org/10.1103/PhysRevLett.99.027404}
  {\bibfield  {journal} {\bibinfo  {journal} {Phys. Rev. Lett.}\ }\textbf
  {\bibinfo {volume} {99}},\ \bibinfo {pages} {027404} (\bibinfo {year}
  {2007})}\BibitemShut {NoStop}%
\bibitem[28]{sm:Hochberg:2021pkt}%
  \BibitemOpen
  \bibfield  {author} {\bibinfo {author} {\bibfnamefont {Y.}~\bibnamefont
  {Hochberg}}, \bibinfo {author} {\bibfnamefont {Y.}~\bibnamefont {Kahn}},
  \bibinfo {author} {\bibfnamefont {N.}~\bibnamefont {Kurinsky}}, \bibinfo
  {author} {\bibfnamefont {B.~V.}\ \bibnamefont {Lehmann}}, \bibinfo {author}
  {\bibfnamefont {T.~C.}\ \bibnamefont {Yu}},\ and\ \bibinfo {author}
  {\bibfnamefont {K.~K.}\ \bibnamefont {Berggren}},\ }\bibfield  {title}
  {\bibinfo {title} {{Determining Dark-Matter\textendash{}Electron Scattering
  Rates from the Dielectric Function}},\ }\href
  {https://doi.org/10.1103/PhysRevLett.127.151802} {\bibfield  {journal}
  {\bibinfo  {journal} {Phys. Rev. Lett.}\ }\textbf {\bibinfo {volume} {127}},\
  \bibinfo {pages} {151802} (\bibinfo {year} {2021})},\ \Eprint
  {https://arxiv.org/abs/2101.08263} {arXiv:2101.08263 [hep-ph]} \BibitemShut
  {NoStop}%
\bibitem[33]{sm:gpaw}%
  \BibitemOpen
  \bibfield  {author} {\bibinfo {author} {\bibfnamefont {J.~J.}\ \bibnamefont
  {Mortensen}}, \bibinfo {author} {\bibfnamefont {A.~H.}\ \bibnamefont
  {Larsen}}, \bibinfo {author} {\bibfnamefont {M.}~\bibnamefont {Kuisma}},
  \bibinfo {author} {\bibfnamefont {A.~V.}\ \bibnamefont {Ivanov}}, \bibinfo
  {author} {\bibfnamefont {A.}~\bibnamefont {Taghizadeh}}, \bibinfo {author}
  {\bibfnamefont {A.}~\bibnamefont {Peterson}}, \emph {et~al.},\ }\bibfield
  {title} {\bibinfo {title} {{GPAW}: An open {Python} package for electronic
  structure calculations},\ }\href {https://doi.org/10.1063/5.0182685}
  {\bibfield  {journal} {\bibinfo  {journal} {The Journal of Chemical Physics}\
  }\textbf {\bibinfo {volume} {160}},\ \bibinfo {pages} {092503} (\bibinfo
  {year} {2024})}\BibitemShut {NoStop}%
\bibitem[34]{sm:yan2011}%
  \BibitemOpen
  \bibfield  {author} {\bibinfo {author} {\bibfnamefont {J.}~\bibnamefont
  {Yan}}, \bibinfo {author} {\bibfnamefont {J.~J.}\ \bibnamefont {Mortensen}},
  \bibinfo {author} {\bibfnamefont {K.~W.}\ \bibnamefont {Jacobsen}},\ and\
  \bibinfo {author} {\bibfnamefont {K.~S.}\ \bibnamefont {Thygesen}},\
  }\bibfield  {title} {\bibinfo {title} {Linear density response function in
  the projector augmented wave method: Applications to solids, surfaces, and
  interfaces},\ }\href {https://doi.org/10.1103/PhysRevB.83.245122} {\bibfield
  {journal} {\bibinfo  {journal} {Phys. Rev. B}\ }\textbf {\bibinfo {volume}
  {83}},\ \bibinfo {pages} {245122} (\bibinfo {year} {2011})}\BibitemShut
  {NoStop}%
\bibitem[35]{sm:dressel2002electrodynamics}%
  \BibitemOpen
  \bibfield  {author} {\bibinfo {author} {\bibfnamefont {M.}~\bibnamefont
  {Dressel}}\ and\ \bibinfo {author} {\bibfnamefont {G.}~\bibnamefont
  {Gr{\"u}ner}},\ }\href {https://books.google.com/books?id=Wv-wrtRdWT0C}
  {\emph {\bibinfo {title} {Electrodynamics of Solids: Optical Properties of
  Electrons in Matter}}}\ (\bibinfo  {publisher} {Cambridge University Press},\
  \bibinfo {address} {Cambridge, England},\ \bibinfo {year} {2002})\BibitemShut
  {NoStop}%
\bibitem[49]{sm:Griffin:2025wew}%
  \BibitemOpen
  \bibfield  {author} {\bibinfo {author} {\bibfnamefont {S.~M.}\ \bibnamefont
  {Griffin}}, \bibinfo {author} {\bibfnamefont {Y.}~\bibnamefont {Hochberg}},
  \bibinfo {author} {\bibfnamefont {B.~V.}\ \bibnamefont {Lehmann}}, \bibinfo
  {author} {\bibfnamefont {R.}~\bibnamefont {Ovadia}}, \bibinfo {author}
  {\bibfnamefont {K.~A.}\ \bibnamefont {Persson}}, \bibinfo {author}
  {\bibfnamefont {B.~A.}\ \bibnamefont {Suter}}, \bibinfo {author}
  {\bibfnamefont {R.~X.}\ \bibnamefont {Yang}},\ and\ \bibinfo {author}
  {\bibfnamefont {W.}~\bibnamefont {Zhao}},\ }\bibfield  {title} {\bibinfo
  {title} {{First High-Throughput Evaluation of Dark Matter Detector
  Materials}},\ }\href {https://doi.org/10.1103/bb4g-7939} {\bibfield
  {journal} {\bibinfo  {journal} {Phys. Rev. Lett.}\ }\textbf {\bibinfo
  {volume} {136}},\ \bibinfo {pages} {191801} (\bibinfo {year}
  {2026})}\BibitemShut {NoStop}%
\bibitem[50]{sm:DAMIC:2019dcn}%
  \BibitemOpen
  \bibfield  {author} {\bibinfo {author} {\bibfnamefont {A.}~\bibnamefont
  {Aguilar-Arevalo}} \emph {et~al.} (\bibinfo {collaboration} {DAMIC Collaboration}),\
  }\bibfield  {title} {\bibinfo {title} {{Constraints on Light Dark Matter
  Particles Interacting with Electrons from DAMIC at SNOLAB}},\ }\href
  {https://doi.org/10.1103/PhysRevLett.123.181802} {\bibfield  {journal}
  {\bibinfo  {journal} {Phys. Rev. Lett.}\ }\textbf {\bibinfo {volume} {123}},\
  \bibinfo {pages} {181802} (\bibinfo {year} {2019})},\ \Eprint
  {https://arxiv.org/abs/1907.12628} {arXiv:1907.12628 [astro-ph.CO]}
  \BibitemShut {NoStop}%
\bibitem[51]{sm:DAMIC-M:2023gxo}%
  \BibitemOpen
  \bibfield  {author} {\bibinfo {author} {\bibfnamefont {I.}~\bibnamefont
  {Arnquist}} \emph {et~al.} (\bibinfo {collaboration} {DAMIC-M Collaboration}),\ }\bibfield
  {title} {\bibinfo {title} {{First Constraints from DAMIC-M on Sub-GeV
  Dark-Matter Particles Interacting with Electrons}},\ }\href
  {https://doi.org/10.1103/PhysRevLett.130.171003} {\bibfield  {journal}
  {\bibinfo  {journal} {Phys. Rev. Lett.}\ }\textbf {\bibinfo {volume} {130}},\
  \bibinfo {pages} {171003} (\bibinfo {year} {2023})},\ \Eprint
  {https://arxiv.org/abs/2302.02372} {arXiv:2302.02372 [hep-ex]} \BibitemShut
  {NoStop}%
\bibitem[52]{sm:SENSEI:2023zdf}%
  \BibitemOpen
  \bibfield  {author} {\bibinfo {author} {\bibfnamefont {P.}~\bibnamefont
  {Adari}} \emph {et~al.} (\bibinfo {collaboration} {SENSEI Collaboration}),\ }\bibfield
  {title} {\bibinfo {title} {{First Direct-Detection Results on Sub-GeV Dark
  Matter Using the SENSEI Detector at SNOLAB}},\ }\href
  {https://doi.org/10.1103/PhysRevLett.134.011804} {\bibfield  {journal}
  {\bibinfo  {journal} {Phys. Rev. Lett.}\ }\textbf {\bibinfo {volume} {134}},\
  \bibinfo {pages} {011804} (\bibinfo {year} {2025})},\ \Eprint
  {https://arxiv.org/abs/2312.13342} {arXiv:2312.13342 [astro-ph.CO]}
  \BibitemShut {NoStop}%
\bibitem[61]{sm:JMLR:v6:banerjee05a}%
  \BibitemOpen
  \bibfield  {author} {\bibinfo {author} {\bibfnamefont {A.}~\bibnamefont
  {Banerjee}}, \bibinfo {author} {\bibfnamefont {I.~S.}\ \bibnamefont
  {Dhillon}}, \bibinfo {author} {\bibfnamefont {J.}~\bibnamefont {Ghosh}},\
  and\ \bibinfo {author} {\bibfnamefont {S.}~\bibnamefont {Sra}},\ }\bibfield
  {title} {\bibinfo {title} {Clustering on the unit hypersphere using von
  Mises--Fisher distributions},\ }\href
  {https://www.jmlr.org/papers/v6/banerjee05a.html} {\bibfield  {journal}
  {\bibinfo  {journal} {Journal of Machine Learning Research}\ }\textbf
  {\bibinfo {volume} {6}},\ \bibinfo {pages} {1345} (\bibinfo {year}
  {2005})}\BibitemShut {NoStop}%
\end{thebibliography}

\newpage
%%%%%%%%%%%%%%%%%%%%%%%%%%%%%%%%%
\onecolumngrid
\section*{End Matter}
\begin{figure}[h]
    \includegraphics[width=0.9\linewidth,trim=0 180 0 40,clip]{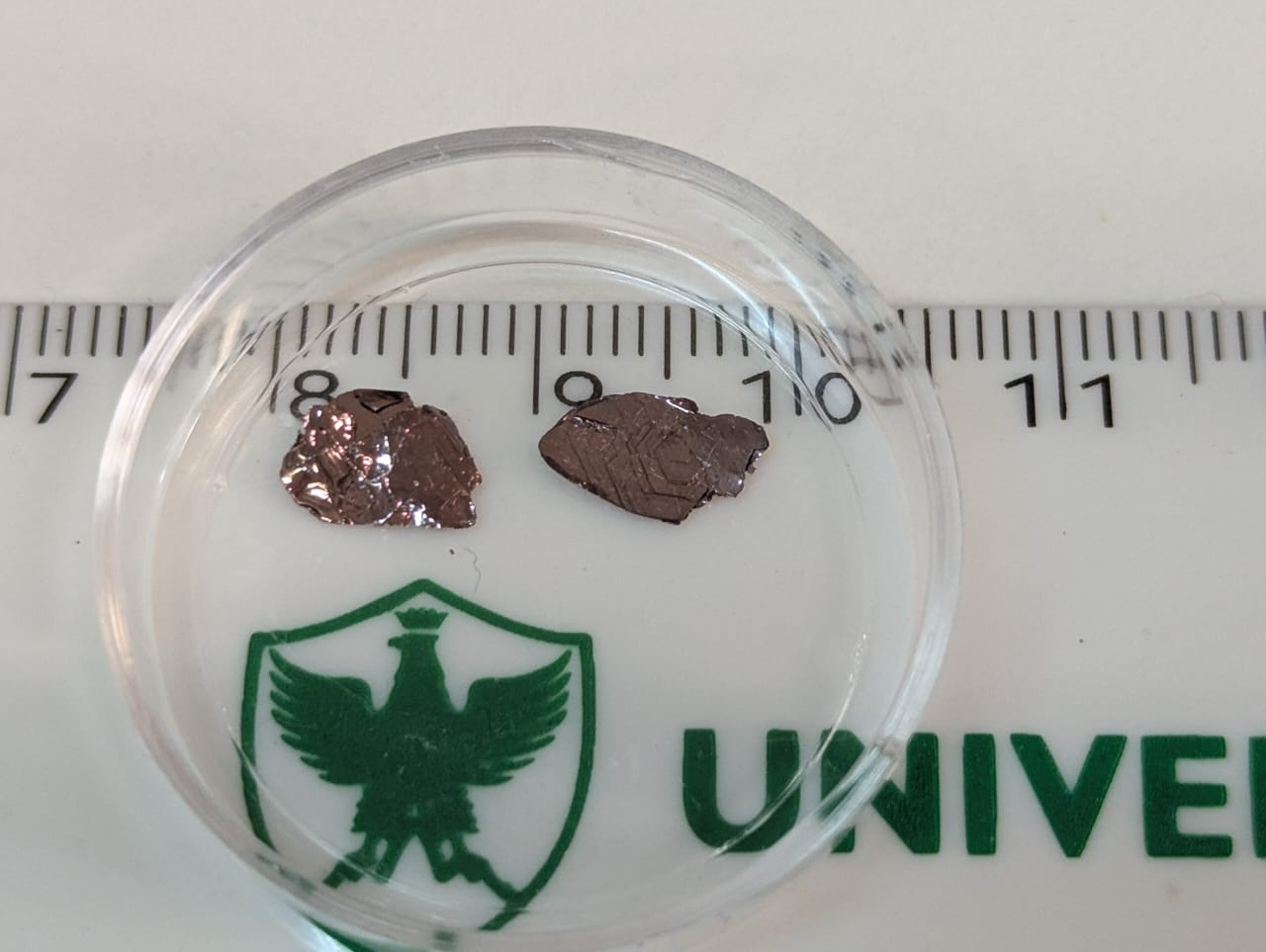}
    \caption{{\bf Synthesized TiSe$_2$ samples.}
    The two TiSe$_2$ milligram-scale flakes synthesized using chemical vapor transport in the Politano lab at L'Aquila.}
    \label{fig:flakes}
\end{figure}

\twocolumngrid
%%%%%%%%%%%%%%%%%%%%%%%%%%%%%%%%%
%
{\it Material synthesis}---All three materials considered in this Letter have been experimentally studied and are synthesizable in laboratory quantities \cite{Li:2007, Kogar2017, lin2022, husain2023, bhattacharya2025}.
To illustrate this, we briefly describe the scalable synthesis of TiSe$_2$ crystals using chemical vapor transport,
following well-established procedures.
TiSe$_2$ can be synthesized from elemental Ti and Se in sealed quartz ampules placed in a two-zone furnace.
References~\cite{Hotje:2005xxxx,Ubaldini:2014xxxx,Wang:2016xxxx,Yang:2025xxxx} report gram-scale batches per ampule, with growth times of one to two weeks and stoichiometry controlled at the level of $\lesssim 1\%$ in Se content (this level of purity is necessary for preserving the CDW in the material response).
We have performed an additional experimental test of these protocols at the Politano lab at L'Aquila using a single-ampule, one-week chemical vapor transport run. 
The run produced two millimeter-scale TiSe$_2$ crystals with a total mass of a few tens of milligrams, shown in Fig.~\ref{fig:flakes}, demonstrating that the growth protocol is both simple and robust.
With an optimized setup utilizing a few tens of ampules in parallel, production rates of hundreds of grams per month are expected to be achievable at a total operational cost below US \$1000 per kilogram.

{\it Experimental considerations}---Beyond material synthesis, a full detector concept based on the targets presented in this Letter will require coupling the material  to a suitable low-threshold readout. 
The sub-eV plasmonic excitations produced by a DM interaction are expected to decay into electronic and phononic excitations, which may be measured calorimetrically or through nonequilibrium charge collection at electrical contacts either at the surface or embedded in the material. 
A full demonstration of sensitivity in the lowest-energy region will also require dedicated sensor optimization.  
The same low thresholds that make these materials sensitive to light DM also require addressing thermal and infrared photons, radiogenic and surface backgrounds, and cosmogenic secondaries---motivating cryogenic operation, shielding, and ideally underground deployment.
Crucially, the anisotropic response of the targets we propose serves as a strong discriminator between a DM signal and isotropic or stationary backgrounds.

{\it Data handling}---In order to obtain the full anisotropic response of the target materials, we take the directionally dependent DFT results of $\epsilon_n(\omega,q)=\vu{q}_n\cdot\overset{\scriptscriptstyle\leftrightarrow}{\epsilon}(\omega,\vb{q}\parallel\vu{q}_n)\cdot\vu{q}_n$ for the directions $\vu{q}_n$ we have computed, and utilize the crystal symmetries to determine all directions with equivalent responses, and consequently interpolate over the unit sphere using a von Mises-Fisher kernel~\cite{JMLR:v6:banerjee05a}.
This procedure ensures our loss function respects all crystal symmetries and aligns with data at all equivalent directions to $\vu{q}_n$. 
We describe the interpolation scheme in more detail in SM~\cite{supplemental}.

{\it Statistical analysis}---Regarding DM scattering with electrons, our analysis considers both the detection prospects for an isotropic signal, as well as the projected ability to detect a directional daily modulation signal.
For the former, we compute the projected reach at 95\% CL of each target, assuming Poisson statistics and no background~\cite{Feldman:1997qc} ({\it i.e.}, a three-event reach). We compare the daily-averaged scattering rate along each of the Cartesian directions and select the direction yielding the best experimental reach.

To determine the prospects to detect an anisotropic signal, we perform a two-bin analysis (AM vs PM) for each material to determine the significance of its respective daily modulation signals. 
The significance is calculated via Monte Carlo, where the DM-target scattering events are assumed to be distributed according to a Poisson distribution, with the different rates in AM and PM accounting for the daily modulation, and no directional backgrounds are assumed.
Through this procedure, we discern the minimal number of scattering events required to observe a difference between the two bins at $95\%$ CL, which we denote as $N_{\rm directional}$. 
Since $N_{\rm directional}$ depends on the rate modulation, it varies between the different materials and model parameters ({\it e.g.}, $m_\chi$, $m_{\phi}$). 
We estimate the reach of experiments relying on a daily modulating signal to discern a DM signal using the $N_{\rm directional}$-event reach.
Similarly to the isotropic case, we compare the $N_{\rm directional}$-event reach along each of the Cartesian directions, selecting the one with the best experimental prospects. 
Note that this yields a conservative reach since the required $N_{\rm directional}$ events would be smaller for a more elaborate statistical test utilizing, {\it e.g.}, a multibin~analysis.

For the case of absorption, we take the loss function in Eq.~\eqref{eq:isotropic-absorption} to be the angle-averaged loss function 
$\widehat{\cal W}(\omega, q) = \frac{1}{4\pi} \int d\Omega_{\vb{q}} \, {\cal W}(\omega, \vb{q})$,
using the zero momentum limit, in order to establish the prospects of each material to probe the parameter space of kinetically mixed dark photon DM.

%%%%%%%%%% Supplemental Material %%%%%%%%%%

\onecolumngrid{}
\clearpage

\setcounter{page}{1}
\setcounter{equation}{0}
\setcounter{figure}{0}
\setcounter{table}{0}
\setcounter{section}{0}
\setcounter{subsection}{0}
\renewcommand{\theequation}{S.\arabic{equation}}
\renewcommand{\thefigure}{S\arabic{figure}}
\renewcommand{\thetable}{S\arabic{table}}
\renewcommand{\thesection}{\Roman{section}}
\renewcommand{\thesubsection}{\Alph{subsection}}
\renewcommand{\theHequation}{supplement.equation.\arabic{equation}}
\renewcommand{\theHfigure}{supplement.figure.\arabic{figure}}
\renewcommand{\theHtable}{supplement.table.\arabic{table}}
\renewcommand{\theHsection}{supplement.section.\arabic{section}}
\renewcommand{\theHsubsection}{supplement.subsection.\arabic{section}.\arabic{subsection}}
\newcommand{\ssection}[1]{
    \addtocounter{section}{1}
    \oldsection{\thesection.~~~#1}
    \addtocounter{section}{-1}
    \refstepcounter{section}
    \noindent\ignorespaces{}
}
\newcommand{\ssubsection}[1]{
    \addtocounter{subsection}{1}
    \subsection{\thesubsection.~~~#1}
    \addtocounter{subsection}{-1}
    \refstepcounter{subsection}
    \noindent\ignorespaces{}
}
\newcommand{\fakeaffil}[2]{$^{#1}$\textit{#2}\\}

\thispagestyle{empty}
\begin{center}
    \begin{spacing}{1.2}
        \textbf{\large 
            Supplemental Material:\\
            Unconventional Materials for Light Dark Matter Detection
        }
    \end{spacing}
    \par\smallskip
    Yonit Hochberg,\textsuperscript{1,2}
    Dino Novko,\textsuperscript{3}
    Rotem Ovadia,\textsuperscript{1,2}
    and Antonio Politano\textsuperscript{4}
    \par\smallskip
    {\small
        \fakeaffil{1}{Racah Institute of Physics, Hebrew University of Jerusalem, Jerusalem 91904, Israel}
        \fakeaffil{2}{Laboratory for Elementary Particle Physics,
 Cornell University, Ithaca, NY 14853, USA}
        \fakeaffil{3}{Centre for Advanced Laser Techniques, Institute of Physics, Zagreb 10000, Croatia}
        \fakeaffil{4}{Department of Physical and Chemical Sciences, University of L’Aquila, L’Aquila 67100, Italy}
        (Dated: \today)
    }

\end{center}
\par\smallskip

In this Supplemental Material, we give additional details of our density functional theory (DFT) dielectric tensor data; our anisotropic fitting procedure; a comparison of DFT results versus optical data in TiSe$_2$; and the role of acoustic plasmons in the reach of hole-doped diamond (HDD). The numbering of references matches that of the main text.

\begin{figure}
    \centering
    \includegraphics[width=0.99\linewidth]{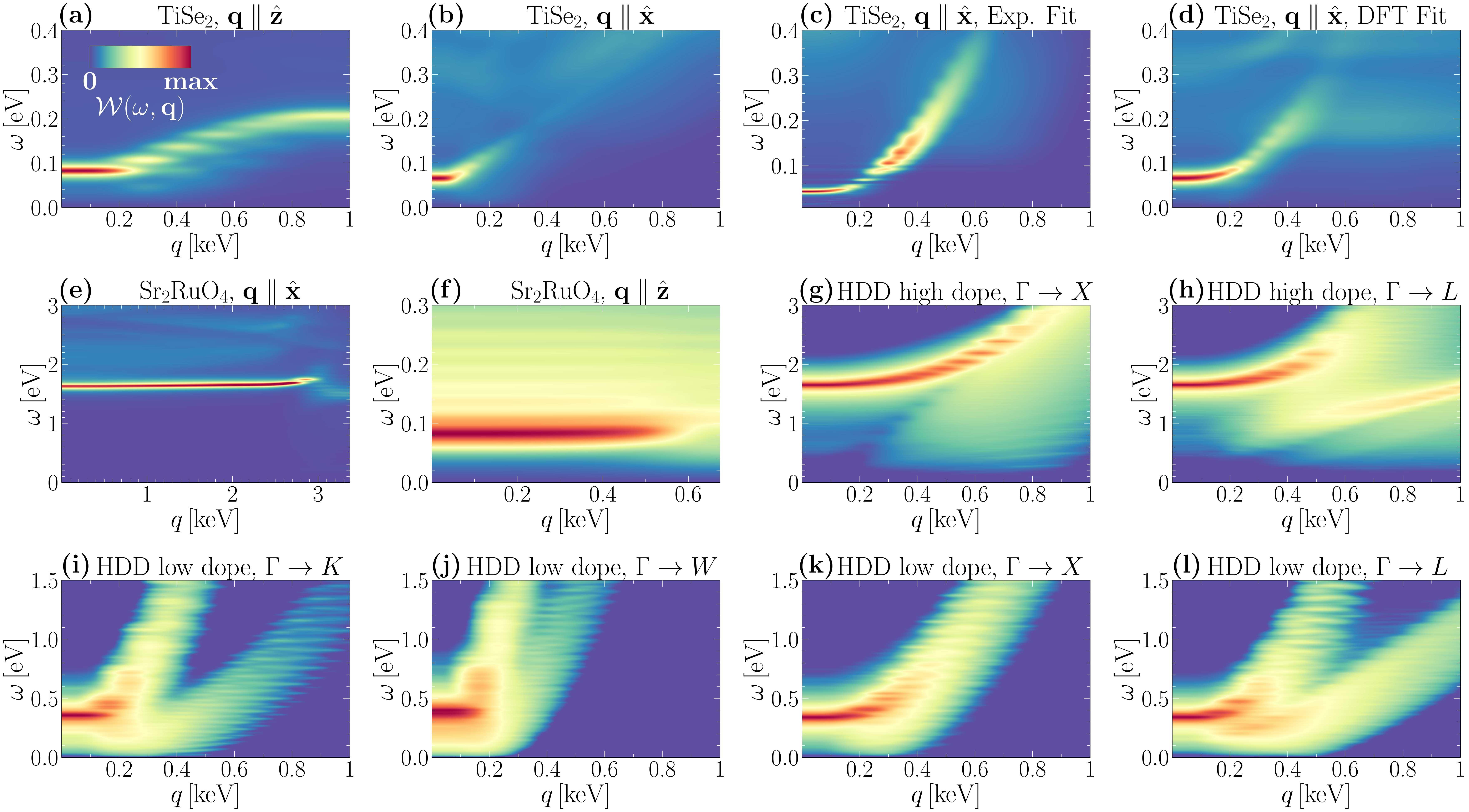}
    \caption{{\bf Loss Functions.} A comprehensive overview of all dielectric loss functions used in this work. Momenta are given in units where $1\,{\rm keV} = 0.507\,$\AA$^{-1}$.
    Panels {\bf (a)} and {\bf (b)} show the DFT response of TiSe$_2$ at $T=30\, {\rm K}$ along the $\vu{z}$ and $\vu{x}$ directions.
    Panels {\bf (c)} and {\bf (d)} show fits to $q = 0$ data in TiSe$_2$---one using experimental optical measurements at $T = 10\,{\rm K}$ in the $xy$ plane, and the other from DFT calculations in the $\vu{x}$ direction---using the fit procedure of Eq.~\eqref{eq:loss fit}.
    Panels {\bf (e)} and {\bf (f)} show full DFT calculations for Sr$_2$RuO$_4$ along the $\vu{x}$ and $\vu{z}$ directions, respectively.
    Panels {\bf (g)} and {\bf (h)} present the response of HDD with high doping $n_h = 4.52 \times 10^{21} \, {\rm cm}^{-3}$ along the $\Gamma \to X$ and $\Gamma \to L$ directions.
    Panels {\bf (i)–(l)} display the same for HDD with low doping $n_h = 4.52 \times 10^{19} \, {\rm cm}^{-3}$ in the $\Gamma \to K$, $\Gamma \to W$, $\Gamma \to X$, and $\Gamma \to L$ directions. 
    } 
    \label{fig:all loss function}
\end{figure}

%%%%%%%%%%%%%%%%%%%%%%%%%%%%%%%%%%%%%%%%%%%%%%%%%%%
\ssection{DFT Data}
%%%%%%%%%%%%%%%%%%%%%%%%%%%%%%%%%%%%%%%%%%%%%%%%%%%
%
In this work, we performed DFT computations to construct the loss function of our materials of interest at finite momenta and energy.
The DFT calculations of the ground state, needed for the loss functions, were performed with the GPAW package~\cite{gpaw}. The plane-wave basis set with the energy cutoff of 45\,Ry and Perdew–Burke–Ernzerhof~(PBE) exchange-correlation functional were used in all of the cases. In order to relax the interlayer distances in TiSe$_2$ and Sr$_2$RuO$_4$ we additionally used dispersion corrections with the vdW-DF functional. The unit cell parameter and atomic positions were relaxed until the forces were below $10^{-6}$\,Ry/$a_0$, with $a_0$ the Bohr radius. 
The unit cell parameters were $a=3.537$\,\AA~(in-plane), $a=3.922$\,\AA~(in-plane), and  $a=3.536$\,\AA~for TiSe$_2$, Sr$_2$RuO$_4$, and HDD, respectively. The out-of-plane cell size were $c=6.309$\,\AA~ and $c=13.03$\,\AA~ for TiSe$_2$ and Sr$_2$RuO$_4$. 
To simulate the TiSe$_2$ in the low-temperature charge-density wave (CDW) structural phase, we use $2\times 2\times 1$ cell with periodic lattice distortions along the eigenvectors of CDW amplitude phonon mode. The CDW supercell is then relaxed at low electron temperatures $T_e=100$\,K (namely below the CDW transition temperature), {\it i.e.}, low electron smearing in Fermi-Dirac functions.
The momentum grids for ground-state density calculations were $\mathbf{k}=18\times 18\times 6$, $\mathbf{k}=6\times 6\times 2$, and $\mathbf{k}=24\times 24\times 24$ for TiSe$_2$, and Sr$_2$RuO$_4$, and HDD, respectively. The denser grids needed for the corresponding loss function calculations~\cite{yan2011} were $\mathbf{k}=50\times 50\times 15$, $\mathbf{k}=120\times 120\times 20$ and $\mathbf{k}=90\times 90\times 90$, respectively.
We present the comprehensive collection of DFT computed loss functions we use in this manuscript in Figs.~\ref{fig:all loss function}{\bf (a),(b),(e)-(l)}.
We note that our computations require finite momentum values. 
As the loss function becomes independent of $q$ at small momenta, the zero-momentum response is approximated by the lowest-momentum points in each grid.

%%%%%%%%%%%%%%%%%%%%%%%%%%%%%%%%%%%%%%%%%%%%%%%%%%%%%%%%%%%%%%%%%%%%%
%%%%%%%%%%%%%%%%%%%%%%%%%%%%%%%%%%%%%%%%%%%%%%%%%%%%%%%%%%%%%%%%%%%%%

%%%%%%%%%%%%%%%%%%%%%%%%%%%%%%%%%%%%%%%%%%%%%%%%%%%%%%%%%%%%%%%%%%
\ssection{Anisotropic Fitting Procedure}
%%%%%%%%%%%%%%%%%%%%%%%%%%%%%%%%%%%%%%%%%%%%%%%%%%%%%%%%%%%%%%%%%%
%
From the directionally dependent DFT calculations of $\epsilon_n(\omega, q) = \vu{q}_n \cdot \epsilon(\omega, \vb{q} \parallel \vu{q}_n) \cdot \vu{q}_n$, performed along several crystal directions $\vu{q}_n$, we reconstruct the full dielectric tensor $\epsilon(\omega, \vb{q})$ as a function of energy and momentum.
To exploit the crystal symmetry, we generate all directions $\vu{q}_{nm}$ 
that are symmetry-equivalent to $\vu{q}_n$, for which the directional response satisfies $\epsilon_{nm}(\omega, q) = \epsilon_n(\omega, q)$ by construction.
We then perform an angular interpolation over these directions using a von Mises–Fisher kernel \cite{JMLR:v6:banerjee05a}:
\begin{eqnarray}
\epsilon(\omega, \vb{q}) = \frac{1}{\cal N} \sum_{nm} \epsilon_{nm}(\omega, q) \exp{\kappa \, \vu{q} \cdot \vu{q}_{nm}}
\end{eqnarray}
with normalization ${\cal N} = \sum_{nm} \exp{\kappa \, \vu{q} \cdot \vu{q}_{nm}}$ and $\kappa = 30$.
This choice of $\kappa$ ensures that $\epsilon(\omega, \vb{q} \parallel \vu{q}_{nm})$ approximately reproduces the input data $\epsilon_{n}(\omega, q)$  
while providing a smooth interpolation between them.
Setting $\kappa=30$ yields a smooth, nearest‐neighbor‐like proxy for the true anisotropic loss, accurate at the ${\cal O}(0.1-1)\, {\rm sr}$ level.
The angular resolution can be improved by sampling additional directions, with the runtime scaling linearly with the number of directions.

%%%%%%%%%%%%%%%%%%%%%%%%%%%%%%%%%%%%%%%%%%%%%%%%%%%
\ssection{DFT vs. Fitting Optical Data}
%%%%%%%%%%%%%%%%%%%%%%%%%%%%%%%%%%%%%%%%%%%%%%%%%%%
%
Here we provide further details related to our TiSe$_2$  
results, where in addition to using our full DFT computation we also use fitted optical data, in order to estimate uncertainties in the DFT-based calculations and 
to assess this material's expected reach into light dark matter (DM) parameter space.

Optical measurements provide access to the zero-momentum dielectric response, $\epsilon_L(\omega, \vb{q}=0)$ of a material, through photon absorption.
They are typically experimentally simpler than electron-electron scattering measurements for extracting material properties, and are relatively abundant. 
We begin by describing our procedure for extrapolating the momentum dependence from optical data, following the method introduced in Ref.~\cite{Griffin:2025wew}.
We construct a momentum-dependent fit based on the features observed in the $\bb q \to 0$ data. 
We identify all energies $\hat\omega_k$ where $\Re(\epsilon_L(\omega, \vb{0}))$ crosses zero. 
All such zeros beyond the first, $\hat\omega_0$, are approximated to correspond to bulk plasmon peaks in the loss function. 
Each of these peaks is fitted using a Lindhard dielectric function $\epsilon_{\rm Lind}(\omega_{k}, \Gamma_{k}, E_{\rm F}; q, \omega)$, where we fix the plasma frequency $\omega_{k} = \hat\omega_k$ and fit the corresponding width $\Gamma_{k}$. 
The expression for the Lindhard dielectric function is taken from Refs.~\cite{dressel2002electrodynamics,Hochberg:2021pkt}, from which a Lindhard loss function ${\cal W}_{\rm Lind}(\omega_{k}, \Gamma_{k}, E_F; \omega, q)$ is constructed.
The full loss function response is then modeled as a weighted sum of the fitted components:
\begin{equation}
{\cal W}_{\rm fit} (\omega, q) =
\frac{1}{\sum_k h_k} \sum_{k=1}^{n_{\mathrm{peaks}}}
h_k {\cal W}_{\rm Lind}(\omega_{k}, \Gamma_{k}, E_F; \omega, q) \, ,
\end{equation}
with weights $h_k$ determined from the best fit to the peak heights and 
\begin{eqnarray}
    E_F = \pqty{\frac{9 \pi^2 \, \omega_p}{128 \, \alpha^2 \, m_e}}^{1/3} \omega_p \, .
\end{eqnarray}
To reproduce the original dielectric function in the $\bb q \to 0$ limit, we introduce a momentum-independent correction factor. 
Specifically, we define a residual scaling function
\begin{eqnarray}
    r(\omega) \equiv \frac{{\cal W}_{\rm data}(\omega, \vb{q} = 0)}{{\cal W}_{\rm fit}(\omega,\vb{q} = 0)} \, , 
\end{eqnarray}
and use it to define a corrected loss function, which we use when presenting `fit' results:
\begin{equation}\label{eq:loss fit}
    \mathcal{W}_r(\bb q, \omega) = r(\omega) \times {\cal W}_{\rm fit}(\omega, \bb q) \, .
\end{equation}
This construction retains the momentum dependence implied by the Lindhard fit while ensuring that the $\bb q \to 0$ limit reproduces the original data.

\begin{figure}[t!]
    \centering
\includegraphics[width=0.975\linewidth]{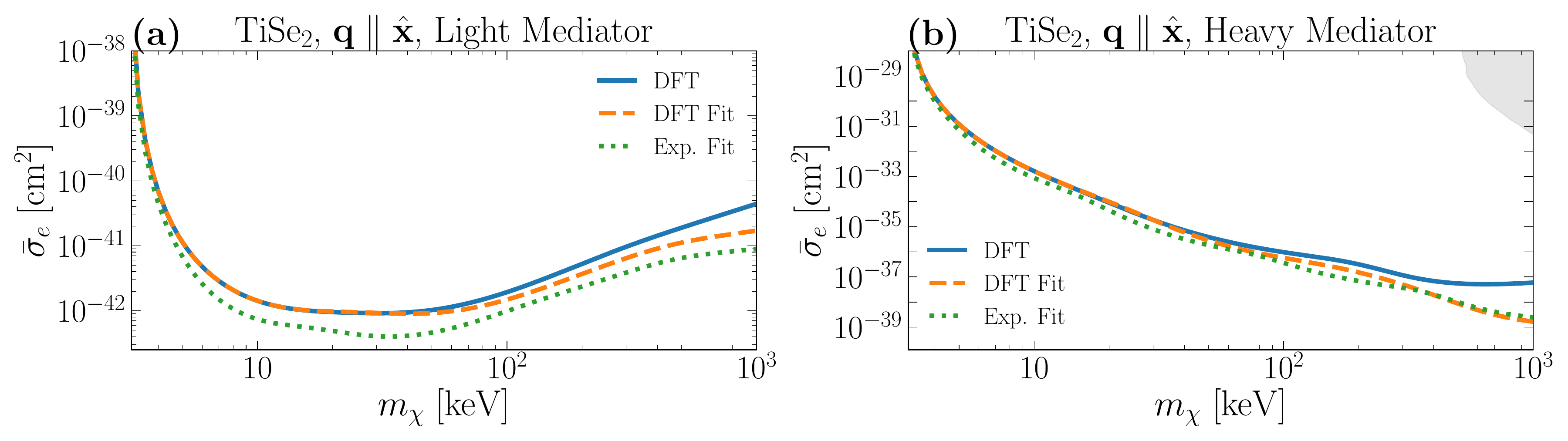}
    \caption{{\bf Fitting method vs. full DFT.} Projected 95\% CL reach for a kg-yr exposure of TiSe$_2$ assuming no background and experimental sensitivity to energy deposits $\omega \in [10\, {\rm meV}, 10 \, {\rm eV}]$, calculated from the loss function of TiSe$_2$ in the $\vu{x}$ direction, assuming a light mediator {\bf (a)} and a heavy mediator {\bf (b)}. 
    The solid blue curve corresponds to the full DFT loss function (illustrated in Fig.~\ref{fig:all loss function}{\bf (b)}), while the orange dot-dashed and green dotted curves represent reach estimates using fitted loss functions based on $q=0$ DFT data (shown in Fig.~\ref{fig:all loss function}{\bf (d)}) and experimental optical data (presented in Fig.~\ref{fig:all loss function}{\bf (c)}), respectively.
    The gray shaded region corresponds to existing terrestrial constraints~\cite{DAMIC:2019dcn,DAMIC-M:2023gxo,SENSEI:2023zdf}.
}
    \label{fig:TiSe2 comparison of fits}
\end{figure}

We emphasize two key assumptions inherent in the fitting procedure: 
{\it (i)}~it assumes that all relevant material excitations are visible in the input $q=0$ data, 
and {\it (ii)}~it assumes a dispersion relation of the form $\omega^2 = \omega_k^2 + q^2$ for the $k_{\rm th}$ peak.
The first assumption is intrinsic to the use of optical data: we cannot model features that are not visible in the measured response. 
Any material excitations not visible in the optical data will necessarily be absent from the fitted model.

To assess the validity of the second assumption we present the case study of TiSe$_2$ in the $\vu{x}$ direction. 
We consider three loss functions representing the response: the DFT calculated loss (Fig.~\ref{fig:all loss function}{\bf (b)}), a fit to the $q \simeq 0$ slice of the DFT loss (Fig.~\ref{fig:all loss function}{\bf(d)}), and a fit to the $q \simeq 0$ loss function as measured at $T=10\,{\rm K}$ via optical spectroscopy in Ref.~\cite{Li:2007} (the $q \simeq 0$ Exp. loss is shown in Fig.~1{\bf (d)}  
and the resulting fit is given in Fig.~\ref{fig:all loss function}{\bf (c)}).
Fig.~\ref{fig:TiSe2 comparison of fits} presents the projected scattering rate of TiSe$_2$ at $95\%$~CL, for a background free kg-yr exposure and an energy detection range $\omega \in [10\, {\rm meV}, 10 \, {\rm eV}]$, as calculated with each of the loss functions in Figs.~\ref{fig:all loss function}{\bf (b)-(d)}.  
Despite qualitative differences between the three loss functions, we find ${\cal O}(1)$ agreement in the projected scattering rates.
Furthermore, at sub-hundred keV masses, we find agreement between the scattering rate as calculated with the full DFT loss function and the momentum-dependent fit using only its $q=0$ slice at the $10\%$ level.
We note that the same test was applied to the rest of the DFT-calculated loss functions Figs.~\ref{fig:all loss function}{\bf (a),(b),(e)$-$(l)}, where we find that the fitted DFT data  
induces only ${\cal O}(1)$ differences in the scattering rate compared to the full DFT data, 
throughout, thereby establishing the validity of the fitting procedure 
of Ref.~\cite{Griffin:2025wew}
for the energy range considered~here.

Note that the experiment fit curves in Fig.~\ref{fig:TiSe2 comparison of fits} trace the outline of the blue‑shaded region in Figs.~2\textbf{(a)–(b)}. The DFT curves in Fig.~\ref{fig:TiSe2 comparison of fits}, however, do not coincide with the solid blue lines in Figs.~2\textbf{(a)–(b)} because Fig.~\ref{fig:TiSe2 comparison of fits} assumes the isotropic response of Fig.~\ref{fig:all loss function}\textbf{(b)} for all momenta directions, whereas Figs.~2\textbf{(a)–(b)} use an anisotropic model incorporating the responses in Figs.~\ref{fig:all loss function}\textbf{(a)} and \textbf{(b)}.

\newpage
%%%%%%%%%%%%%%%%%%%%%%%%%%%%%%%%%%%%%%
\ssection{Acoustic Plasmons}
%%%%%%%%%%%%%%%%%%%%%%%%%%%%%%%%%%%%%%
%
\begin{figure}[t]
    \centering
    \includegraphics[width=0.975\linewidth]{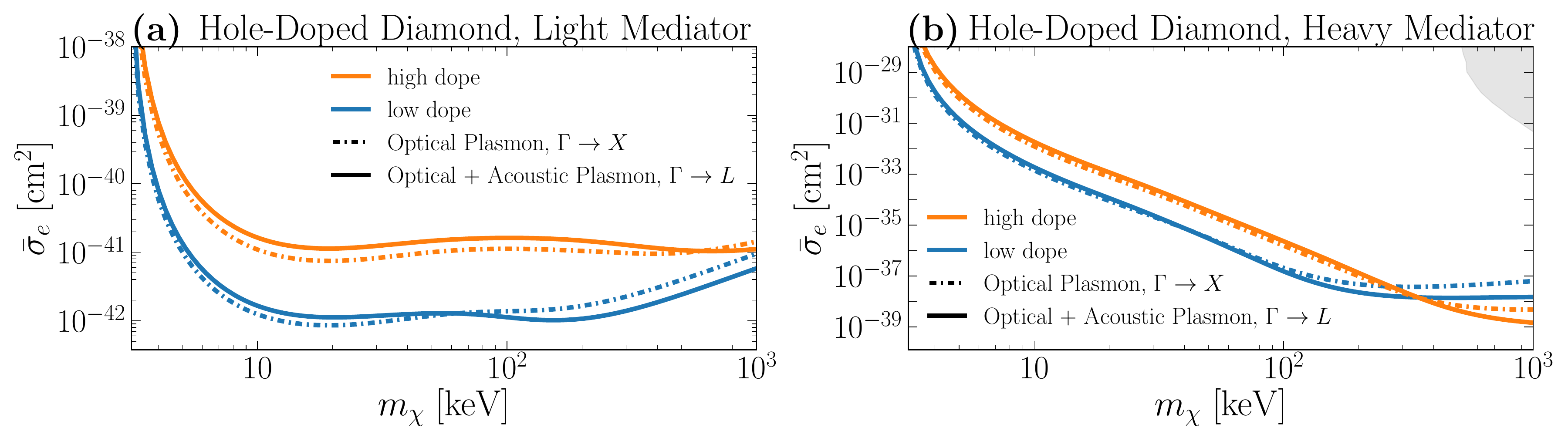}
    \caption{{\bf Acoustic plasmons in HDD.} The projected reach using the response of HDD along two different crystal directions $\Gamma \to X$~(dot-dashed) and $\Gamma \to L$~(solid) for high doping (orange) and low doping (blue), assuming a light mediator {\bf (a)} and a heavy
mediator {\bf (b)}. 
    The $\Gamma \to X$ direction only has an optical plasmon, whereas the $\Gamma \to L$ direction also has an acoustic plasmon. 
    The high (low) dope curves correspond to hole doping of $n_h = 4.52 \times 10^{21} \, {\rm cm}^{-3}$ ($4.52 \times 10^{19} \, {\rm cm}^{-3}$).
    The gray shaded region corresponds to existing terrestrial constraints~\cite{DAMIC:2019dcn,DAMIC-M:2023gxo,SENSEI:2023zdf}.
    }
    \label{fig:acoustic plasmons}
\end{figure}

HDD exhibits both optical and acoustic plasmon modes at low energies. 
The acoustic plasmon is absent along the $\Gamma \to X$ direction but emerges along $\Gamma \to L$, as illustrated in Figs.~\ref{fig:all loss function}{\bf (g)$-$(h)} and Figs.~\ref{fig:all loss function}{\bf (k)$-$(l)}.
In Fig.~\ref{fig:acoustic plasmons}, we compare the projected scattering reach in HDD computed independently using the dielectric response in each of these crystallographic directions, for low and high dopings.
We find that for both dopings, the two directions yield comparable sensitivity despite the broader, but weaker, support over high-momenta provided by the acoustic plasmon along $\Gamma \to L$.

%%%%%%%%%% Supplemental references %%%%%%%%%%

\begingroup
\makeatletter
\let\sm@originallabel\label
\def\label#1{%
	\def\sm@testlabel{#1}%
	\def\sm@lastbiblabel{LastBibItem}%
	\ifx\sm@testlabel\sm@lastbiblabel
		\sm@originallabel{LastSMBibItem}%
	\else
		\sm@originallabel{#1}%
	\fi
}

\makeatother
\endgroup

\end{document}